\documentclass{style/nseJournal}
\usepackage{amsmath}
\usepackage{setspace}
\usepackage{amssymb}
\usepackage{epsfig}
\usepackage{amscd}
\usepackage{subcaption}
\usepackage{float}
\usepackage[normalem]{ulem}

\usepackage{bm}
\usepackage[dvipsnames]{xcolor}
\usepackage{siunitx}
\usepackage{algorithm, algorithmic}
\usepackage[numbers]{natbib}

\begin{document}

\title{The effect of branchless collisions and population control on correlations in Monte Carlo power iteration}

\addAuthor{{Theophile Bonnet}}{a}
\addAuthor{Hunter Belanger}{b}
\addAuthor{\correspondingAuthor{Davide Mancusi}}{a}
\correspondingEmail{davide.mancusi@cea.fr}
\addAuthor{Andrea Zoia}{a}
\addAffiliation{a}{Universite Paris-Saclay, CEA, \\Service d'Etudes des Reacteurs et de Mathematiques Appliquees, \\91191, Gif-sur-Yvette, France}%
\addAffiliation{b}{Department of Mechanical, Aerospace and Nuclear Engineering,\\ Rensselaer Polytechnic Institute,\\ Troy, NY 12180, USA}%

\addKeyword{Monte Carlo simulations}
\addKeyword{Spatial correlations}
\addKeyword{Population control}
\addKeyword{Power iteration}
\addKeyword{Branchless collisions}

\titlePage

\begin{abstract}
The investigation of correlations in Monte Carlo power iteration has been long dominated by the question of generational correlations and their effects on the estimation of statistical uncertainties. More recently, there has been a growing interest in spatial correlations, prompted by the discovery of neutron clustering. Despite several attempts, a comprehensive framework concerning how Monte Carlo sampling strategies, population control and variance reduction methods affect the strength of such correlations is still lacking. In this work, we propose a set of global and local (i.e., space-dependent) tallies that can be used to characterize the impact of correlations. These tallies encompass the Shannon entropy, the pair distance, the normalized variance and the Feynman moment. In order to have a clean, yet fully meaningful setting, we carry out our analysis in a few homogeneous and heterogeneous benchmark problems of varying dominance ratio. Several classes of collision sampling strategies, population control and variance reduction techniques are tested, and their relative advantages and drawbacks are assessed with respect to the proposed tallies. The major finding of our study is that branchless collisions, which suppress the emergence of branches in neutron histories, also considerably reduce the effects of correlations in most of the explored configurations.
\end{abstract}

\section{Introduction}

In the context of reactor physics or nuclear criticality safety, the $k$-eigenvalue formulation of the Boltzmann equation is often tackled by using the well-known Monte Carlo power iteration method, which relies on the iteration of the fission source: neutrons starting a given `generation' are chosen amongst the neutrons born from fission in the previous generation~\cite{bell_nuclear_1970,lux_monte_1991}. After a sufficient number of generations, the phase-space distribution of fission neutrons eventually becomes stationary, and so do the related tallies, such as neutron flux and reaction rates. Once the fission source has converged to a stationary state, the sought tallies are averaged using an ergodic average over successive generations in order to reduce the statistical uncertainty on the estimated values.

For the purpose of estimating confidence intervals on the ergodic averages, most production-level Monte Carlo codes for the sake of simplicity ignore correlations between successive generations, although statistical tests are becoming progressively available, fostered by the increased awareness of Monte Carlo practitioners. Ground-breaking investigations concerning the impact of correlations on the estimation of variance in Monte Carlo power iteration have been carried out by Brissenden and Garlick, who noted that inference techniques based on assuming independent and identically distributed samples lead to an underestimation of the sample variance of the average multiplication factor $k_{\text{eff}}$~\cite{BRISSENDEN198663}. The severity of the underestimation was later related to the value of the dominance ratio (i.e.\ the ratio of the second over the first eigenvalue of the Boltzmann equation) by Ueki et al.~\cite{ueki_autocorrelation_2003}, and since then several authors have tried to predict and correct the estimate~\cite{herman_monte_2014,miao_analysis_2016,sutton_application_2017,miao_predicting_2018,miao_correlation_2019}.

More recently, Dumonteil et al.\ have pointed out that fission-induced correlations may lead to a strong spatial patchiness (i.e.\ non-Poisson spatial fluctuations) in the fission source, which was dubbed \textit{neutron clustering} \cite{dumonteil_particle_2014}. A characterization of spatial correlations relying on moment equations was derived in a time-dependent context in simplified nuclear reactor models~\cite{zoia_clustering_2014,mulatier_critical_2015,zoia_neutron_2017,bonnet_space_2022}, and the key findings were showed to apply also to power iteration~\cite{nowak_monte_2016,sutton_neutron_2017} and to production-level Monte Carlo simulations of realistic configurations~\cite{dumonteil_patchy_2021}. Several methods have been proposed in order to detect and possibly quench the effects of clustering. Nowak et al.\ suggested to use Shannon entropy for neutron clustering detection~\cite{nowak_monte_2016}, by pointing out that entropy is systematically decreased by the presence of neutron clustering. Inspired by pioneering work in population dynamics by Zhang et al.\ \cite{zhang_diffusion_1990}, Sutton introduced the concept of neutron lineage in power iteration and outlined the relation between neutron clustering and fixation (i.e.\ the number of generations at which all neutrons in the Monte Carlo simulation share the same common ancestor)~\cite{sutton_neutron_2017,sutton_toward_2022}. Equations for the moments of the fixation time in kinetic simulations have been derived~\cite{bonnet_fixation_2023}, which might be used to establish a more rigorous theoretical framework for fixation in power iteration.

Over the last few years, the issue of correlations, and spatial correlations in particular, has led to a collection of somewhat contradictory results. Cosgrove et al.\ pointed out the impact of clustering, suggesting that it might induce spurious xenon oscillations in depletion calculations coupling Monte Carlo neutronics solvers and Bateman solvers \cite{cosgrove_neutron_2020}. Fr\"ohlicher et al.\ suggested that neutron clustering could lead to an apparent bias on the average Monte Carlo tallies, in addition to affecting the estimation of the variance \cite{frohlicher_generational_2023}. In contrast, Mickus and Dufek stated that neutron clustering should not be detrimental to criticality Monte Carlo simulations, in that the estimate of the variance is asymptotically not altered \cite{mickus_does_2021}.

In view of the lack of a consensus on these matters, which are highly relevant for both conceptual and practical questions related to power iteration, we have decided to explore in some depth the emergence of spatial correlations and how they are affected by the key mechanisms at play in Monte Carlo simulations, namely the collision sampling strategies and the population control and variance reduction methods. For this purpose, in this work we propose a set of global and local (i.e., space-dependent) tallies that can be used to characterize the impact of correlations. These tallies encompass the Shannon entropy, the pair distance, the normalized variance and the Feynman moment. In order to have a clean, yet fully meaningful setting, we carry out our analysis in a few homogeneous and heterogeneous benchmark problems of varying dominance ratio. The relative advantages and drawbacks of the collision sampling strategies, population control and variance reduction techniques are assessed with respect to the proposed tallies. We focus in particular on the use of branchless collisions, which has been independently introduced in the context of criticality by Fr\"ohlicher et al.\ in Ref.~\citenum{frohlicher_generational_2023} and Belanger et al.\ in Ref.~\citenum{belanger_clustering_2023}.

This paper is organized as follows. In Section~\ref{sec:model} we recall the main ideas behind the Monte Carlo power iteration. The relation between \textit{generational} and \textit{spatial} correlations is made explicit, and we present the different variance reduction techniques used here. Next, in Section~\ref{sec:geometry}, we describe a set of benchmark problems, and we introduce the tallies that we propose to investigate spatial correlations. Section~\ref{sec:pi} is devoted to the analysis of numerical results, and we provide conclusions in Section~\ref{sec:conclusions}.

\section{Power iteration in Monte Carlo simulation}
\label{sec:model}

The dynamics of neutrons in multiplying systems at position ($\overrightarrow{r}$), direction ($\overrightarrow{\Omega}$) and energy ($E$) is described by the Boltzmann equation. In order to characterize the deviation of the system from the stationary regime, it is customary to introduce the $k$-eigenvalue formulation of the Boltzmann equation, namely:
\begin{align}
    \overrightarrow{\Omega} \cdot \nabla\phi_k(\overrightarrow{r}, \overrightarrow{\Omega},E) +& \Sigma_t(\overrightarrow{r}, E) \phi_k(\overrightarrow{r},\overrightarrow{\Omega},E) = \iint\Sigma_s(\overrightarrow{r}, \overrightarrow{\Omega'} \rightarrow \overrightarrow{\Omega}, E' \rightarrow E) \phi_k(\overrightarrow{r}, \overrightarrow{\Omega}', E') d\overrightarrow{\Omega'}dE' \nonumber \\
    & + \frac{1}{k}\frac{1}{4\pi}\iint_{4\pi}\chi(\overrightarrow{r},E' \rightarrow E)\nu(\overrightarrow{r}, E') \Sigma_f(\overrightarrow{r}, E') \phi_k(\overrightarrow{r}, \overrightarrow{\Omega'}, E') \, d\overrightarrow{\Omega'} \, dE'\text,
    \label{eq:static_boltzmann}
\end{align}
where $k$ is the eigenvalue and $\phi_k$ the corresponding eigenfunction. The notation is standard: $\Sigma_t$ is the macroscopic total cross section, $\Sigma_s$ is the  macroscopic double-differential scattering cross section, $\Sigma_f$ is the macroscopic fission cross-section, $\nu$ is the average fission yield, and $\chi$ the fission emission spectrum. The dominant eigenpair $\lbrace k_0, \phi_{k_0} \rbrace$ of Eq.~\eqref{eq:static_boltzmann} can be estimated by either deterministic or Monte Carlo methods: Monte Carlo simulation is to be preferred to deterministic solvers for the solution of Eq.~\eqref{eq:static_boltzmann} whenever reference results free of discretization errors are sought~\cite{lux_monte_1991}. For this purpose, the standard approach relies on a stochastic version of the well-known power iteration method~\cite{lux_monte_1991}. Starting from an arbitrary tentative fission source, a `neutron generation' is sampled: each history of the initial particle population consists of a succession of exponential flights interrupted by collisions, whereupon neutrons undergo scattering, fission, or capture. The neutrons are followed until they are absorbed (by fission or capture) or leak out of the system. Upon fission, on average, $\nu$ descendant neutrons are sampled and stored in a `fission bank'. When all the histories of a given generation terminate, the neutrons in the fission bank are promoted to source neutrons for the next generation. This new generation of source neutrons will undergo the same transport process, producing a new fission bank which can be used as the source for a subsequent generation. After a sufficiently large number of such iterations, the neutrons starting all subsequent generations are distributed according to the fundamental mode for the fission source. Correspondingly, the neutron flux within each generation converges towards the fundamental eigenmode $\phi_{k_0}$, and the ratio between the statistical weights of neutrons in two successive fission sources converges towards the fundamental eigenvalue $k_0$. The convergence of the power iteration from an arbitrary initial condition to the stationary state is primarily driven by the \textit{dominance ratio} $R$, i.e.\ the ratio $k_1/k_0$~\cite{bell_nuclear_1970}. The dominance ratio describes how fast higher-order eigenmodes decay compared to the fundamental eigenmode, and is thus related to the number of neutron generations that are needed to attain convergence. These generations are usually called inactive, since one is not allowed to collect useful statistics for the sought observables during this phase of the power iteration.

As apparent from the above description, neutrons belonging to the same fission chain within the power iteration algorithm are correlated, and these correlations are carried over successive generations. Correlations manifest themselves in two ways: a neutron in generation $g$ can be spatially correlated with other neutrons in the same generation, which is referred to as \textit{spatial correlations}; furthermore, a neutron in generation $g$ can be correlated with another neutron at generation $g+k$, which is referred to as \textit{generational correlations}. Due to the fact that correlations arise from the branching process underlying neutron dynamics (i.e.\ fission), this distinction is somewhat artificial: spatial correlations result from generational correlations, and vice versa, so that the two points of view are complementary. The behavior of correlations is intimately related to the dominance ratio: as $R$ approaches $1$, the spatial distribution may become patchy (neutron clustering) and strongly non-Poissonian fluctuations appear. It has been shown that the dominance ratio affects the rate at which neutrons `forget' the fact of sharing a common ancestor in previous generations~\cite{ueki_autocorrelation_2003}.

\subsection{Variance reduction and population control methods}
\label{sec:varred}

In order to prevent the population size from growing out of control or dying out over the course of many generations, the total statistical weight of the fission source is kept constant. The statistical dispersion of neutron weights can be further reduced by using \textit{optional} population control algorithms: \textit{i)} local population control may be applied in the course of a generation on individual neutrons (e.g.\ Russian roulette and splitting are applied to neutrons emerging from collision events~\cite{lux_monte_1991}), and \textit{ii)} global population control can be applied to the fission bank, before neutrons are promoted to source neutrons for the following generation (e.g.\ weight combing is applied to fission neutrons in order to select source neutrons~\cite{booth_weight_1996}).

The sampling of neutron histories can be either analog or non-analog; in the latter case, neutrons are assigned weight correction factors to keep the Monte Carlo game unbiased. In this work, the sampling of the flight kernel will always be analog, whereas for collisions we will use various non-analog sampling methods, in order to probe their effectiveness as variance reduction techniques. We shall use implicit capture and forced fission, which are customary in most production Monte Carlo codes~\cite{brun2014tripoli,MCNP6,leppanen2015serpent,romano_openMC_2015}. The number of fission neutrons will be sampled via
\begin{equation}
    n_f = \left\lfloor \xi + \frac{\nu(\overrightarrow{r},E) \Sigma_f(\overrightarrow{r},E)}{\Sigma_{t}(\overrightarrow{r},E)}\right\rfloor,
\end{equation}
where $\xi \sim {\cal U}[0,1]$ and $\lfloor \boldsymbol{\cdot} \rfloor$ denotes the integer part; fission neutrons are assigned the weight $w$ of their parent. The parent neutron is forced to undergo scattering, and the effect of absorption is taken into account by correcting the statistical weight by the survival probability, namely,
\begin{equation}
    w' = w \frac{\Sigma_s(\overrightarrow{r},E)}{\Sigma_t(\overrightarrow{r},E)}.
\end{equation}
In this non-analog scheme, particles never disappear by absorption; histories can be terminated by either leakage or Russian roulette after collision events~\cite{lux_monte_1991}.

The sampling strategy described so far belongs to the class of `branching' algorithms, in the sense that new particles (hence new branches of the particle histories) can be created at collision events through the sampling of fission progeny. Alternatively, a `branchless' algorithm can be applied, ensuring that a single neutron is sampled at collisions~\cite{lux_monte_1991}. Branchless collisions have been promoted by Sjenitzer and Hoogenboom in the context of kinetic simulations~\cite{sjenitzer_monte_2011,sjenitzer_dynamic_2013}, and recently extended by Belanger et al.\ and Fr\"ohlicher et al.\ to $k$-eigenvalue calculations~\cite{frohlicher_generational_2023,belanger_clustering_2023}. Following Belanger et al., at each collision event the scattering or fission reaction channel is chosen with probability
\begin{equation}
 P_s = \frac{\Sigma_s(\overrightarrow{r},E)}{\Sigma_s(\overrightarrow{r},E) + \nu(\overrightarrow{r},E) \Sigma_f(\overrightarrow{r},E)}  \qquad \text{and} \qquad
 P_f = \frac{\nu(\overrightarrow{r},E) \Sigma_f(\overrightarrow{r},E)}{\Sigma_s(\overrightarrow{r},E) + \nu(\overrightarrow{r},E) \Sigma_f(\overrightarrow{r},E)},
\end{equation}
respectively. In either case, a single outgoing neutron is sampled, and is assigned a new weight
\begin{equation}
    w' = w \frac{\Sigma_s(\overrightarrow{r},E) + \nu(\overrightarrow{r},E) \Sigma_f(\overrightarrow{r},E)}{\Sigma_t(\overrightarrow{r},E)}.
\end{equation}
If scattering is chosen, energy and direction are sampled as customary, and the emitted neutron performs a new flight within the same history. If fission is chosen, the emitted neutron is sent to the fission bank and the current history is terminated: each neutron undergoes at most one fission event during a generation, producing at most one descendant per generation, whence the name `branchless'\footnote{Observe that branchless collisions reduce to branching collisions with implicit capture in regions without fission.}. Within this sampling strategy, population control is mandatory: the number of neutrons would otherwise decrease over generations, since neutrons disappearing by leakage or Russian roulette would not be replaced. In practice, applying splitting to particles emerging from collision events suffices to counteract this limitation, provided that the number of particles per generation is sufficiently large. 

Other population control mechanisms can be applied to power iteration. One such method is combing, originally proposed by Booth \cite{booth_weight_1996} and recently explored at length in the context of time-dependent simulations~\cite{faucher_new_2018,variansyah_analysis_2022}. The idea behind combing is essentially to enforce a constant number of particles in the source bank of each generation while preserving the total weight. The source particles are sampled using a constant weight interval over the total weight, starting with a random offset which ensures unbiasedness. An alternative to combing relies on the sampling without replacement (WOR) method, proposed by Sutton \cite{sutton_toward_2022}. Using sampling WOR, source neutrons for the next generation are sampled without replacement from the list of weighted fission sites. Although the implementation of sampling WOR described in Ref.~\citenum{sutton_toward_2022} applies to a simplified quasi-analog model which excludes Roulette and splitting, the generalization to a more realistic framework including non-analog particle transport is straightforward: the source neutrons are sampled without replacement from the weighted distribution of fission neutrons. Each sampled neutron is assigned the statistical weight of the fission neutron it originates from. Since sampling WOR is an improvement of the Duplicate-Discard strategy presented by Variansyah et McLarren~\cite{variansyah_analysis_2022}, in the following we focus exclusively on sampling WOR.

\subsection{Estimating the correlations of the sought observables}

Most production simulation codes perform \textit{ergodic} averages of the sought observables over successive `active' generations, after convergence has been attained, and compute an \textit{apparent} variance assuming (for ease of computation) that successive generations are independent. In practice, generational correlations cause the apparent variance to underestimate the \textit{real} variance of the stochastic process. Several methods have been developed to recover the real variance from the apparent variance, for instance by block-averaging tallies over several generations in order to wash out correlations, or by modeling generational correlations in the system and applying suitable corrections to the apparent variance \cite{demaret1999accurate,sutton_application_2017,miao_analysis_2016,miao_predicting_2018}. Alternatively, one might take \textit{ensemble} averages over fully independent power iteration replicas: in this case, the real variance of the stochastic process would be estimated, but this would come at the cost of an often intolerable increase in simulation time.

By construction, the collision sampling strategies and the variance reduction and population control methods described in Sec.~\ref{sec:varred} do not affect the estimate of the average, since these algorithms are precisely chosen to ensure a fair (i.e.\ unbiased) Monte Carlo game. Although Fr\"ohlicher et al.\ suggested that spatial clustering may induce a bias in the estimation of the neutron flux, the number of inactive generations they had chosen for their system appears to be insufficient to ensure proper convergence to the fundamental eigenstate~\cite{frohlicher_generational_2023}. In spite of the unbiasedness of the variance-reduction techniques, the estimates of the critical flux and effective multiplication value in power-iteration calculations are affected by a well-known statistical bias, which is due to weight normalization at the beginning of each generation~\cite{BRISSENDEN198663}. To first order, the bias is inversely proportional to the number of particles per generation. In Fr\"ohlicher et al.'s case, the number of particles is arguably so small ($N=10^3$) that the normalization bias cannot be excluded. Moreover, the bias does not necessarily manifest itself in the same way for branchless or branching collisions. Therefore, the bias observed in Ref.~\citenum{frohlicher_generational_2023} cannot be confidently attributed to spatial clustering.

In any case, the non-analog methods presented above do have an impact on the generational and spatial correlations of the resulting simulation scheme, and therefore on the apparent variance of the sought observables. The goal of this work is to characterize the effects of collision sampling strategies, variance reduction and population control on correlations, using several kinds of {\it ad hoc} estimators, encompassing those that are commonly met for power iteration and some new ones that are hopefully more easily amenable to the interpretation of the simulation results. Five algorithms (i.e.\ combinations of collision sampling strategies and population control algorithms) will be considered: \textit{i)} branching collisions; \textit{ii)} branching collisions with sampling WOR; \textit{iii)} branchless collisions; \textit{iv)} branchless collisions with weight combing; \textit{v)} branchless collisions with sampling WOR. Russian roulette and splitting are always applied to the particles emerging from collision events. Note that when no population control algorithm is specified, the fission source is wholly transmitted to the next generation. Part of the present work builds upon a preliminary investigation carried out in Ref.~\citenum{belanger_clustering_2023} for branchless collisions.

\begin{table}[t]
    \centering
        \begin{tabular}{cccccc}
        \hline
        \multirow{2}{*}{Number} & \multirow{2}{*}{Type} & Inactive & Lengths & Dominance & $k_{\text{eff}}$ \\
         &  & generations & ($\unit{cm}$) & ratio &\\
        \hline
        1 & Homogeneous & $100$ & 50 & $0.9832$ & $1.00003 \pm 1~\unit{pcm}$ \\
        2 & Homogeneous & $200$ & 100 & $0.9957$ & $0.99986 \pm 1~\unit{pcm}$ \\
        3 & Heterogeneous & 20 & $20/2/20$ & $0.7908$ & $0.99999\pm 3~\unit{pcm}$\\
        4 & Heterogeneous & 1000 & $20/20/20$ & $0.9971$ & $0.99987 \pm 5~\unit{pcm}$ \\
        \hline
        \end{tabular}
        \caption{Parameters for the four benchmark configurations. The dominance ratio and the value of the capture cross-sections are taken from Vitali et al.~\cite{vitali_eigenvalue_2021}.}
        \label{tab:configurations}
\end{table}

\section{Specifications for a set of benchmark problems}
\label{sec:geometry}

In view of the considerations mentioned in the previous section, we introduce a set of benchmark problems and we provide the key specifications.

\subsection{Description of the chosen configurations}

For the purpose of our analysis, we have chosen a total of four one-dimensional critical slab geometries taken from Vitali et al.~\cite{vitali_eigenvalue_2021}. The selected configurations have different values of dominance ratio $R$, and cover both homogeneous and heterogeneous systems, which are known to behave differently with respect to the parameter $R$. The geometrical data are listed in Tab.~\ref{tab:configurations}. All configurations have leakage boundary conditions at both sides of the slab. The first two configurations (1 and 2 in Tab.~\ref{tab:configurations}) are homogeneous, with different lengths. The material is a mixture of $\text{UO}_2$ and $\text{H}_2\text{O}$, whose parameters are provided in Tab.~\ref{tab:hom_cs}. Configurations 3 and 4 are heterogeneous, with a $\text{UO}_2$-$\text{H}_2\text{O}$-$\text{UO}_2$ sandwich structure, with lengths given in Tab.~\ref{tab:configurations}. The respective nuclear data are given in Tab.~\ref{tab:uo2_p} for the fuel and in Tab.~\ref{tab:h2o} for the moderator. The nuclear data for all the materials are intended to be realistic, and are represented using a three-group formalism, with thermal, epithermal and fast energy groups. In Ref.~\citenum{vitali_eigenvalue_2021}, prompt $\chi_p$ and delayed $\chi^j_d$ spectra for each precursor family $j$ were provided separately. For this work, we collapsed them into an average fission spectrum $\chi$, with $\chi = (\nu_p \chi_p +\sum_j \nu_d^j \chi^j_d) / \nu$.

\begin{table}[t]
    \centering
    \begin{tabular}{cccc}
        \hline
         Parameters & $l=1$ (fast) & $l=2$ (epithermal) & $l=3$ (thermal) \\
         \hline
         $\Sigma_{f,l}$ $[\unit{cm^{-1}}]$ & $3.0586\times \num{e-3}$ & $2.1579\times \num{e-3}$ & $5.6928\times \num{e-2}$ \\
         $\Sigma_{s,1 \rightarrow l}$ $[\unit{cm^{-1}}]$ & $4.48187\times \num{e-1}$ & $1.78483\times \num{e-1}$ & 0 \\
         $\Sigma_{s,2 \rightarrow l}$ $[\unit{cm^{-1}}] $ & $0$ & $1.8058111$ & $1.41764\times \num{e-1}$ \\
         $\Sigma_{s,3 \rightarrow l}$ $[\unit{cm^{-1}}] $ & $0$ & $0$ & $4.31567$ \\
         $\nu_l$ $[-]$ & $2.4$ & $2.4$ & $2.4$ \\
         $\chi_l$ $[-]$ & $0.876304$ & $0.123696$ & 0 \\
         \hline
    \end{tabular}
    \caption{Nuclear data for the homogeneous fuel-moderator material. The three energy groups are indexed by $l$.}
    \label{tab:hom_cs}
\end{table}

\begin{table}[t]
    \centering
    \begin{tabular}{cccc}
        \hline
         Parameters & $l=1$ (fast) & $l=2$ (epithermal) & $l=3$ (thermal) \\
         \hline
         $\Sigma_{f,l}$ $[\unit{cm^{-1}}]  $ & $3.0586\times \num{e-3}$ & $2.1579\times \num{e-3}$ & $5.6928\times \num{e-2}$ \\
         $\Sigma_{s,1 \rightarrow l}$ $[\unit{cm^{-1}}] $ & $2.21062\times \num{e-1}$ & $7.3843\times \num{e-2}$ & 0 \\
         $\Sigma_{s,2 \rightarrow l}$ $[\unit{cm^{-1}}] $ & $0$ & $7.77642\times \num{e-1}$ & $4.3803\times \num{e-2}$ \\
         $\Sigma_{s,3 \rightarrow l}$ $[\unit{cm^{-1}}] $ & $0$ & $0$ & $1.55272$ \\
         $\nu_l$ $[-]$ & $2.4$ & $2.4$ & $2.4$ \\
         $\chi_l$ $[-]$ & $0.876304$ & $0.123696$ & 0 \\
         \hline
    \end{tabular}
    \caption{Nuclear data of the $\text{UO}_2$ material. The three energy groups are indexed by $l$.}
    \label{tab:uo2_p}
\end{table}

\begin{table}[t]
    \centering
    \begin{tabular}{cccc}
    \hline
         Parameters & $l=1$ (fast) & $l=2$ (epithermal) & $l=3$ (thermal) \\
    \hline
        $\Sigma_{c,l}$ $[\unit{cm^{-1}}] $ & $3.05\times \num{e-4}$ &    $3.699\times \num{e-4}$ & $1.825\times \num{e-2}$ \\
        $\Sigma_{s,1 \rightarrow l}$ $[\unit{cm^{-1}}] $ & $2.27125\times \num{e-1}$ & $1.0464\times \num{e-1}$ & 0 \\
        $\Sigma_{s,2 \rightarrow l}$ $[\unit{cm^{-1}}] $ & $0$ & $1.02817$ & $9.7961\times \num{e-2}$ \\
        $\Sigma_{s,3 \rightarrow l}$ $[\unit{cm^{-1}}] $ & $0$ & $0$ & $2.76295$ \\
    \hline
    \end{tabular}
    \caption{Nuclear data for the $\text{H}_2\text{O}$ material. The three energy groups are indexed by $l$.}
    \label{tab:h2o}
\end{table}

\begin{table}[t]
    \centering
    \begin{tabular}{cccc}
    \hline
        System Number & $\Sigma_{c,1} [\unit{cm^{-1}}]$ & $\Sigma_{c,2} [\unit{cm^{-1}}]$ & $\Sigma_{c,3} [\unit{cm^{-1}}]$ \\
        \hline
        1 & $9.3888852 \times \num{e-4}$ & $1.5003548 \times \num{e-2}$ & $6.9125769 \times \num{e-2}$ \\
            2 & $9.3920991 \times \num{e-4}$ & $1.5034394 \times \num{e-2}$ & $7.0670812 \times \num{e-2}$ \\

        3 & $3.8396068 \times \num{e-4}$ & $1.1656103 \times \num{e-2}$ & $4.8808556 \times \num{e-2}$ \\
        4 & $3.8089301 \times \num{e-4}$ & $1.1361681 \times \num{e-2}$ & $3.4061203 \times \num{e-2}$ \\
        \hline
    \end{tabular}
    \caption{Capture section for each configuration. The three energy groups are indexed by $l$.}
    \label{tab:capture_xs}
\end{table}

Following Vitali et al., for all configurations we modify the capture cross section of the homogeneous material (for cases 1 and 2) and of $\text{UO}_2$ (for cases 3 and 4) so that the systems are critical. The system-dependent capture cross-sections are given in Table~\ref{tab:capture_xs}. The dominance ratios for each configuration have been computed by Vitali et al.\ and are also recalled in Tab.~\ref{tab:configurations}.

\subsection{Definition of Estimators for Correlation Diagnostics}
\label{sec:observables}

We introduce a few significant estimators for correlation diagnostics. We require all such estimators to be `non-Boltzmann', i.e.\ they must not be invariant under variance reduction techniques, otherwise they would not be affected by correlations. Their significance in detecting correlations at the global or local spatial scale will be discussed in the following.

\subsubsection{Shannon Entropy}

The Shannon entropy $S$ of the fission source is commonly used as a tool for assessing the convergence of power iteration: the power iteration is assumed to have reached the stationary state when $S$ reaches an asymptotic value $S_\infty$. In view of the fact that the rate of convergence of the Shannon entropy depends on the dominance ratio, and that the dominance ratio also affects the behavior of correlations, the relation between the Shannon entropy of the fission source and the correlation-induced neutron clustering has been examined by Nowak et al.~\cite{nowak_monte_2016}. For this purpose, a Cartesian spatial mesh with $B$ cells is superposed over the system geometry and the Shannon entropy $S$ at generation $g$ is computed as customary:
\begin{equation}
    S(g) = -\sum_{i=1}^B p_i(g) \log_2(p_i(g)), 
    \label{eq:entropy_def}
\end{equation}
where $p_i(g)$ is the fraction of the total weight of the fission source particles in the spatial cell $i$ at generation $g$. In our work, we compute $p_i(g)$ as the ratio of the weight of the fission source particles in cell $i$ over the total weight, after applying normalization. The spatial size of the cells follows from $\delta = L/B$, where $L$ is the linear size of the system. In the simplified homogeneous reactor model tested in  Ref.~\citenum{nowak_monte_2016}, the occurrence of spatial clustering was showed to lead to lower values of the asymptotic Shannon entropy $S_\infty$ compared to the expected theoretical value $S_\textit{id}$ that corresponds to the ideal assumption of $N$ independently and distributed particles of unit weight. The extended analysis performed in Ref.~\citenum{belanger_clustering_2023} shows that these findings apply also to continuous energy.

Observe that the entropy is just a single scalar for each generation: while it does convey information about the presence of spatial clustering, the fine details of the spatial distribution are lost through the coarse-graining in Eq.~\eqref{eq:entropy_def}: only the collective occupation statistics of the cells matters in computing $S(g)$.

Generally speaking, the ideal asymptotic value $S_\textit{id}$ of the entropy function is not known: thus, since the discrepancy between $S_\textit{id}$ and the asymptotic value $S_\infty$ cannot be determined in practice, the value $S(g)$ resulting from the power iteration cannot be directly used to detect the existence of spatial clustering\footnote{To partially overcome this limitation, Nowak et al.\ have proposed to use the spatial moments of the Shannon entropy, weighted by Legendre polynomials along the Cartesian axes~\cite{nowak_monte_2016}: using spatial symmetries occurring in the simulated system, deviations of the moments of the Shannon entropy may help in revealing the emergence of clustering.}. Notwithstanding, the values of Shannon entropy originating from different collision sampling strategies, variance reduction and population control methods can be compared relative to each other, which allows assessing the effectiveness of each technique.

\subsubsection{Pair distance function}

The pair correlation function $C({\bf r}_i,{\bf r}_j,t_i,t_j)$ is a measure of the covariance between the occupation statistics of two spatial sites at positions ${\bf r}_i$ and ${\bf r}_j$, at time $t_1$ and $t_2$, which may coincide. It reads
\begin{equation}
    C({\bf r}_i,{\bf r}_j,t_i,t_j) = \mathbb{E}[n_i(t_i) n_j(t_j)] 
\end{equation}
where $\mathbb{E}[\mathbf{\cdot}]$ denotes the ensemble average, and $n_i(t_i)$ is the occupation number in space bin $i$ at time $t_i$, that may be fractional due to the use of statistical weights. Time can be either continuous (in a time-dependent context) or discrete (i.e.\ generations). In the context of time-dependent diffusion-reproduction stochastic processes in infinite homogeneous media, Meyer et al.\ have shown that the pair correlation function of two spatial sites at a given time characterizes the evolution of particle clusters~\cite{meyer_clustering_1996}. Similar conclusions have been later reached for generalized versions of Meyer's model, encompassing finite-size systems and two observation times~\cite{mulatier_critical_2015,zoia_neutron_2017,bonnet_space_2022}, and also hold true with minor modifications for power iteration, where time $t$ is replaced by discrete generations $g$~\cite{dumonteil_particle_2014,sutton_neutron_2017}. While being highly informative, the pair correlation function may be prohibitively expensive to compute for realistic three-dimensional configurations. A convenient alternative consists in collapsing the information content of the pair correlation function into the mean squared pair distance function (often shortened to pair distance function),
\[
\langle r^2 \rangle(g) \propto \iint |{\bf r}_1-{\bf r}_2|^2 C({\bf r}_1,{\bf r}_2,g)d{\bf r}_1 d{\bf r}_2
\text,
\]
where $C({\bf r}_1,{\bf r}_2,g)$ expresses the correlation function for particle pairs at two positions in the same generation, which provides a tool to estimate the typical cluster size~\cite{meyer_clustering_1996}. The pair distance function is basically the mean squared distance between any two particles, and can be thus straightforwardly computed in Monte Carlo power iteration, e.g., at the beginning of each generation. Since the number of particle pairs scales quadratically with the number of particles present in the generation, the required computational cost might become an issue; however, this problem can be mitigated (without incurring any bias) by estimating the mean squared distance from a maximum number of randomly sampled pairs. Similarly to Shannon entropy, the pair distance function converges towards an asymptotic value $\langle r^2 \rangle_\infty$ for a sufficiently large number of generations, even in the presence of clustering. Departures of $\langle r^2 \rangle_\infty$ from the ideal value, corresponding to a simulation with no clustering, would suggest the presence of spatial clustering. If the distribution is uniform and the particles are independent, then it is easy to estimate the ideal value. In general, however, it is impossible to do so. Similarly to the case of the entropy function, the relative values of the mean squared distance can nonetheless be used to compare the effectiveness of different variance reduction and population control techniques in quenching the effects of correlations.

\subsubsection{The statistics of neutron families}

Neutrons histories simulated in the power iteration algorithm can be assigned an identifier, whose evolution can be monitored along fission generations. In the initial population at the beginning of the power iteration, one starts with $N$ independent neutrons, each of which is assigned a different arbitrary numerical tag identifying an independent `family'. The tags are transmitted to the descendant neutrons at fission events. Two neutrons sharing the same identifier are correlated because they share a common ancestor. During power iteration, the number of neutron families decreases along the generations, because neutrons die and families go extinct by capture or leakage. The intrinsic relationship between diffusion-reproduction processes and particle lineages has been explored by Zhang et al.~\cite{zhang_diffusion_1990} in the context of diffusion-reproduction processes in infinite homogeneous media: they showed that all neutron histories become correlated after a characteristic time proportional to the population size, and they attributed the emergence of particle clustering to this phenomenon. The effects of neutron lineage statistics on the average size and number of clusters has been recently revisited by Sutton et al.~\cite{sutton_neutron_2017,sutton_toward_2022} in the framework of a simplified power iteration model using the WOR sampling method, and in a more realistic multi-group and continuous-energy framework by Belanger et al.~\cite{belanger_clustering_2023} and Fr\"ohlicher et al.~\cite{frohlicher_generational_2023}.

For each independent realization of power iteration, there exists a \textit{fixation generation} at which all but one family have gone extinct: at the fixation generation, all the neutrons are correlated, and the population is said to have reached fixation. All neutrons now share the same ancestor in the initial source. An exact solution for the fixation generation in homogeneous media has been derived in Ref.~\citenum{bonnet_fixation_2023} in the context of kinetic simulations, and these findings extend straightforwardly to processes evolving in discrete generations. We define $\eta(g)$ to be the average number of surviving independent neutron families at generation $g$: the likeliness of occurrence of neutron clustering increases as $\eta$ becomes much smaller than the initial number of families $N$.

While the statistics of neutron families does provide interesting insights about correlations, and for this reason it has been proposed as an estimator for clustering~\cite{sutton_neutron_2017,sutton_toward_2022,belanger_clustering_2023}, the fact that it does not carry information about the spatial and boundary effects may limit its usefulness, as we will show in the following.

\subsubsection{Feynman moments}

The three previous estimators, although sensitive to correlations, are global and do not provide any information about the local distribution of the correlations within the system. As such, their application to the analysis of correlations occurring in systems exhibiting strong spatial heterogeneity is fraught with conceptual difficulties. Ideally, we would like indicators of spatial correlations to be \textit{i)} local, i.e.\ space-dependent, and \textit{ii)} having a known ideal behavior in the absence of correlations, which would make their interpretation easier. For this purpose, we introduce the \textit{ensemble} Feynman moment
\begin{equation}
    Y_{\cal O}({\bf r}_i) = \frac{\mathbb{V}[{\cal O}]({\bf r}_i)}{\mathbb{E}[{\cal O}]({\bf r}_i)},
    \label{eq:defY}
\end{equation}
where ${\bf r}_i$ is the center of the $i$-th spatial cell. Here $\mathbb{E}[\boldsymbol{\cdot}]$ denotes the \textit{ensemble} average over $M$ independent replicas of a given observable ${\cal O}$:
\begin{equation}
    \mathbb{E}[{\cal O}]({\bf r}_i) = \frac{1}{M}\sum_{m=1}^M {\cal O}_{i,m},
\end{equation}
${\cal O}_{i,m}$ being the value of the estimator in cell $i$ for replica $m$. The \textit{ensemble} unbiased estimator of the variance of ${\cal O}$ is defined as
\begin{equation}
    \mathbb{V}[{\cal O}]({\bf r}_i) = \frac{1}{M-1}\sum_{m=1}^M ({\cal O}_{i,m} - \mathbb{E}[{\cal O}]({\bf r}_i))^2,
\end{equation}
We stress that $\mathbb{V}[{\cal O}]$ is the variance of the estimator and not the variance of the mean of the estimator. The variance of the mean is obtained by dividing the variance of the estimator by the number of realizations $M$ over which the mean is performed. The Feynman moment has been originally conceived for the detection of fission-induced correlations in the time series of multiplying systems~\cite{pazsit_neutron_2007}: to the best of our knowledge, its application to the analysis of correlations occurring in power iteration has not been attempted before.

As mentioned in Sec.~\ref{sec:model}, to reduce the cost of power iteration, one generally uses ergodic averages over successive neutron generations. We therefore introduce the \textit{ergodic} estimator of the expectation of ${\cal O}$ over $G$ active generations, namely,
\begin{equation}
    \mathcal{E}[{\cal O}]({\bf r}_i) = \frac{1}{G}\sum_{g=1}^G {\cal O}_{i,g},
    \label{eq:ergodic_av}
\end{equation}
where ${\cal O}_{i,g}$ is the value of the estimator in cell $i$ at active generation $g$. Similarly, one can define the \textit{ergodic} estimator of the variance of ${\cal O}$ by
\begin{equation}
    {\cal V}[{\cal O}]({\bf r}_i) = \frac{1}{G-1}\sum_{g=1}^G ({\cal O}_{i,g} - \mathcal{E}[{\cal O}]({\bf r}_i))^2.
    \label{eq:ergodic_var}
\end{equation}
Because of positive correlations, we expect ${\cal V}[{\cal O}]({\bf r}_i) < \mathbb{V}[{\cal O}]({\bf r}_i) $ when the two estimators are taken over samples of the same size ($G=M$)~\cite{BRISSENDEN198663}. In order to have a better estimate of Eqs.~\eqref{eq:ergodic_av} and \eqref{eq:ergodic_var}, it is possible to perform ensemble averages over $M$ independent replicas on $\mathcal{E}[\boldsymbol{\cdot}]$ and ${\cal V}[\boldsymbol{\cdot}]$. Finally, we define the \textit{ergodic} Feynman moment of ${\cal O}$ by
\begin{equation}
    Y^{\text{G}}_{\cal O}(x) = \frac{\mathbb{E}[{\cal V}[{\cal O}]]({\bf r}_i)}{\mathbb{E}[\mathcal{E}[{\cal O}]] ({\bf r}_i)},
    \label{eq:ergodic_Y}
\end{equation}
which reduces to the ensemble Feynman moment defined in Eq.~\eqref{eq:defY} when $G=1$.

Since the Feynman moment for Poisson variates is equal to $1$, it is possible to interpret its value as the deviation of a given observable (in a spatial cell) from a Poisson-like behavior. A process that generates Poisson-distributed counts is the following: consider $N$ particles, independently and identically distributed into $K$ bins following the probability distribution $p_k$. If $p_k\ll 1$, the number of particles in each bin is approximately Poisson distributed, and the Feynman moment $Y_k$ for bin $k$ is equal to $1$ for all the bins and independently of the probability distribution $p_k$. However, when there are very few bins, then $p_k\sim1$, the counts actually follow a binomial distribution, and we expect $Y < 1$. Finally, suppose that the number of bins is high enough so that the number of particles in each bin is Poisson distributed. The particles are divided into groups of size $\alpha$, and each group is independently tossed into a bin. In this situation, it is easy to check that the Feynman moment is equal to $\alpha$, again for all the bins and independently of the probability distribution. This suggests that the value of the Feynman moment for a counting variable can be interpreted as the typical `weight' of a particle cluster. This example also suggests that Feynman moments are most naturally applied to quantities that are akin to counts. Indeed, for counting variables, the Feynman moment is dimensionless and has a natural reference value to which it can be compared. It is less naturally applicable to commonly used quantities such as fluxes, which is why such quantities are not considered in the present work.

In this work, we apply the Feynman moments to the collision counts $\psi$, i.e.\ the average number of neutrons entering a collision in a space bin\footnote{Given that reaction rates are akin to partial collision counts, one could also analyze their Feynman moments, which will not attempted here, for the sake of conciseness.}, and to the fission emission counts $F$, i.e.\ the average number of fission neutrons being emitted in a space bin. We will focus exclusively on energy-integrated Feynman moments.

\subsubsection{Normalized variance}

Another local estimator for correlations can be introduced using the \emph{ensemble} estimator for the normalized variance
\begin{equation}
    g({\bf r}_i) = \frac{\mathbb{V}[{\cal O}]({\bf r}_i)}{\mathbb{E}[{\cal O}]({\bf r}_i)^2},
\end{equation}
where $\mathbb{E}[{\cal O}]({\bf r}_i)$ and $\mathbb{V}[{\cal O}]({\bf r}_i)$ are defined as above, and ${\bf r}_i$ is again the center of the $i$-th spatial cell. The normalized variance describes the amplitude of fluctuations relative to the average and is a simplified form of the centered normalized pair correlation function that is widely used in the context of spatial clustering analysis~\cite{meyer_clustering_1996,b_houchmandzadeh_neutron_2015,zoia_neutron_2017}. Observe that the normalized variance differs from the Feynman moment only in the normalization terms appearing at the denominator. However, this difference is crucial and conveys a distinct information content with respect to correlations, as illustrated in Sec.~\ref{sec:pi}. The \textit{ergodic} normalized variance $g^{\text{G}}({\bf r}_i)$ can be defined similarly to that of the Feynman moments, namely
\begin{equation}
    g^{\text{G}}({\bf r}_i) = \frac{\mathbb{E}[{\cal V}[{\cal O}]]({\bf r}_i)}{\mathbb{E}[\mathcal{E}[{\cal O}]]({\bf r}_i)^2}.
\end{equation}
In this work, we apply the normalized variance to the same observables as for the Feynman moments, i.e.\ the energy-integrated collision counts and fission emission counts.

\section{Simulation results}
\label{sec:pi}

\begin{figure}[t]
    \centering
    \begin{subfigure}{0.45\textwidth}
        \includegraphics[width=\textwidth]{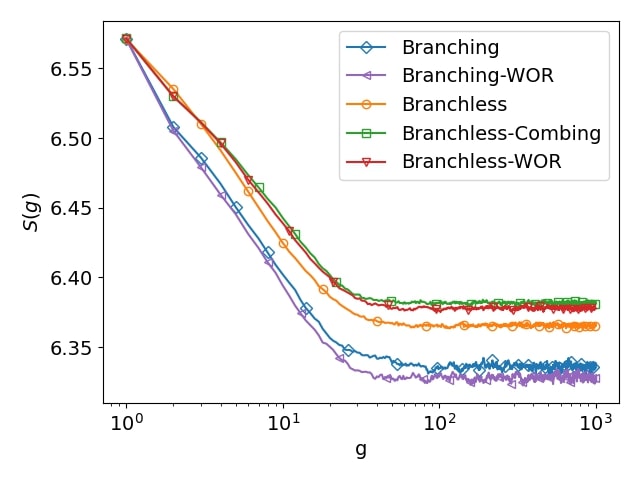}
        \caption{System 1}
        \label{fig:hom_50_S}
    \end{subfigure}
    \begin{subfigure}{0.45\textwidth}
        \includegraphics[width=\textwidth]{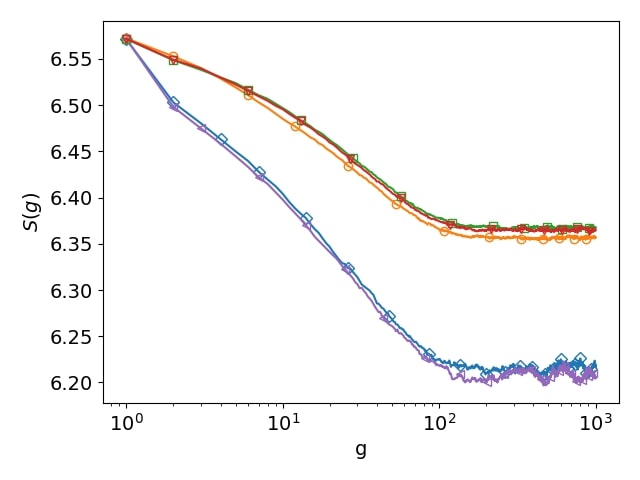}
        \caption{System 2}
        \label{fig:hom_100_S}
    \end{subfigure}
    \begin{subfigure}{0.45\textwidth}
        \includegraphics[width=\textwidth]{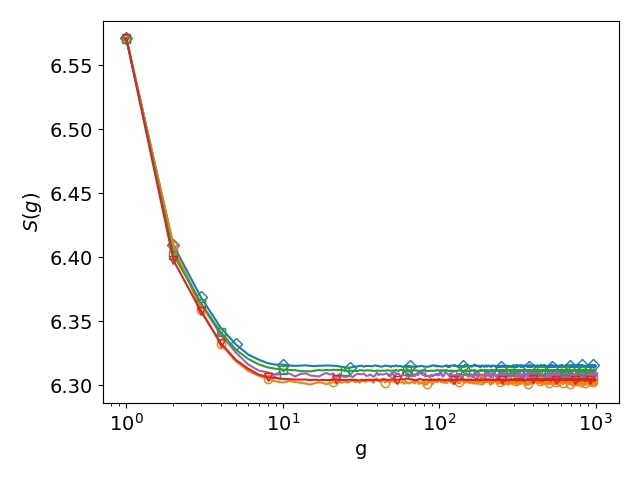}
        \caption{System 3}
        \label{fig:het_42_S}
    \end{subfigure}
    \begin{subfigure}{0.45\textwidth}
        \includegraphics[width=\textwidth]{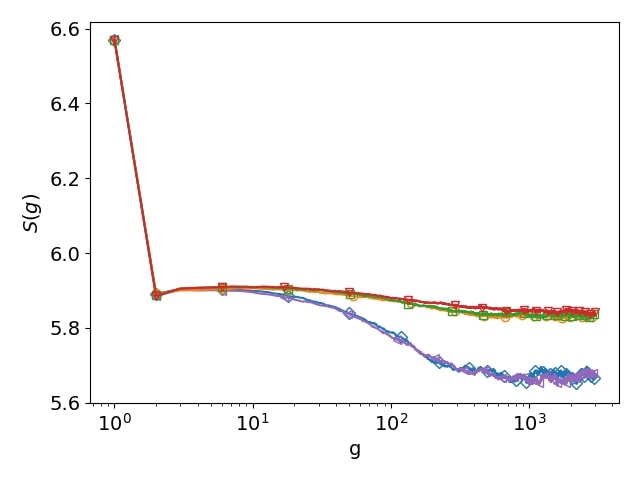}
        \caption{System 4}
        \label{fig:het_60_S}
    \end{subfigure}
    \caption{Shannon entropy as a function of the generations in power iteration, computed for $N=10^3$, $B=10^2$, and for different systems, and averaged over $10^3$ independent replicas.}
    \label{fig:pi_entropy}
\end{figure}

In the following, we present numerical results for the benchmark configurations described in Sec.~\ref{sec:geometry} and the five algorithms presented in Sec.~\ref{sec:model}, first for the global tallies (entropy, pair distance and family number), and then for the local tallies (Feynman moment and normalized variance). The aim of this investigation is to assess the impact of each algorithm on correlations, with the help of the diagnostics tools introduced in Sec.~\ref{sec:observables}. All power iteration simulations carried out for this work have been performed starting from an initial neutron source belonging to the fast group and uniformly distributed in space (non-fissile media included). Russian roulette is assigned a threshold weight $w_R = 0.8$, and splitting is assigned a threshold weight $w_S = 2$. The number of inactive generations for each configuration is given in Tab.~\ref{tab:configurations} and has been determined on the basis of Fig.~\ref{fig:pi_entropy}, while the number of neutrons per generation will be detailed for each case separately.

\subsection{Analysis of Global Estimators}
\label{subsec:global_observables}

\begin{figure}
    \centering
    \begin{subfigure}{0.45\textwidth}
        \includegraphics[width=\textwidth]{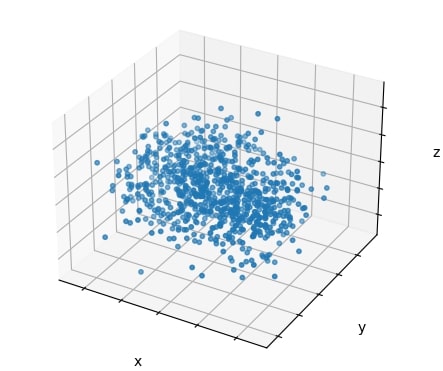}
        \caption{Using branching collisions}
    \end{subfigure}
    \begin{subfigure}{0.45\textwidth}
        \includegraphics[width=\textwidth]{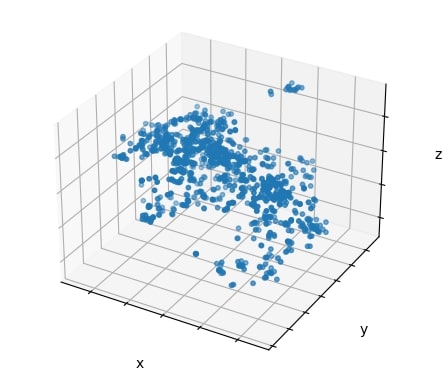}
        \caption{Using branchless collisions}
    \end{subfigure}
    \caption{Snapshots of the positions of neutrons in one realization of system 2, illustrating significant clustering for branching collisions, and almost no clustering for branchless collisions. The distribution of spatial positions is taken after 200 inactive generations to ensure that power iteration has reached the stationary state.}
    \label{fig:clustering}
\end{figure}

We begin our analysis by considering global tallies. The evolution of the Shannon entropy $S(g)$ as a function of the generations in each of the benchmark configurations is illustrated in Fig.~\ref{fig:pi_entropy}. As expected, all entropy curves converge towards an asymptotic value for a large number of generations. The asymptotic value of the entropy depends on the algorithm and the system, whereas the convergence rate (i.e.\ the number of generations needed to attain convergence) only depends on the system. For systems 1, 2, and 4, the asymptotic value of the entropy is smaller for the branching algorithm than for any of the branchless algorithms, suggesting stronger spatial correlations when using branching collisions; this finding is in agreement with previous work~\cite{belanger_clustering_2023}. The typical spatial patchiness induced by spatial correlations is illustrated in Fig.~\ref{fig:clustering} for the case of system 2 (homogeneous reactor with large dominance ratio). It is interesting to note that in system 3 all algorithms lead to similar asymptotic entropy; additionally, the branching algorithm appears to lead to the higher entropy, and is thus expected to induce less spatial correlations. The behavior of system 3 can be probably explained by its small dominance ratio, cf.~Tab.~\ref{tab:configurations}. Spatial clustering is the result of a competition between diffusion and multiplication: correlated branches induce spatial correlations when the neutrons cannot explore the whole system. When using branchless collisions, the only way for correlated branches to appear is through splitting or population control. In systems close to criticality such as ours, the weight correction factor in branchless algorithms is close to unity, which means that splitting seldom occurs during a generation. However, as stated previously, the number of neutrons decreases from one fission bank to the next. Thus, weight normalization increases the average weight of neutrons, ultimately leading to additional splitting. Alternatively, if population control such as combing or sampling WOR is applied, then the same neutron in the fission bank may be copied several times, introducing additional correlated branches.

\begin{figure}[t]
    \centering
    \begin{subfigure}{0.45\textwidth}
        \includegraphics[width=\textwidth]{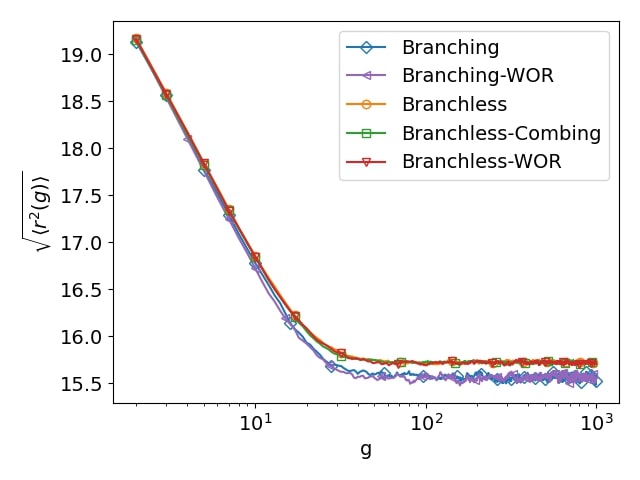}
        \caption{System 1}
        \label{fig:hom_50_R}
    \end{subfigure}
    \begin{subfigure}{0.45\textwidth}
        \includegraphics[width=\textwidth]{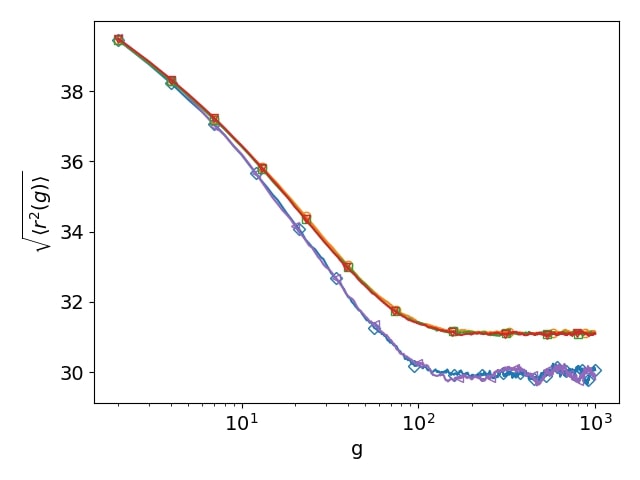}
        \caption{System 2}
        \label{fig:hom_100_R}
    \end{subfigure}
    \begin{subfigure}{0.45\textwidth}
        \includegraphics[width=\textwidth]{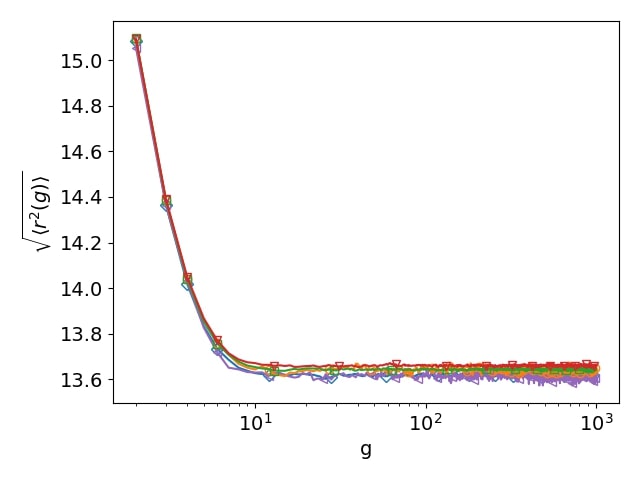}
        \caption{System 3}
        \label{fig:het_42_R}
    \end{subfigure}
    \begin{subfigure}{0.45\textwidth}
        \includegraphics[width=\textwidth]{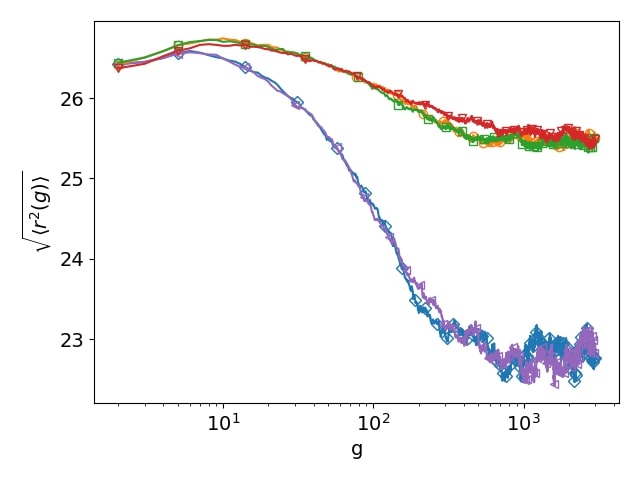}
        \caption{System 4}
        \label{fig:het_60_R}
    \end{subfigure}
    \caption{Square root of pair distance as a function of the generations, computed for $N=10^3$ and $B=10^2$, for different systems, with $10^3$ independent replicas.}
    \label{fig:pi_pair_distance}
\end{figure}

\begin{figure}[t]
    \centering
    \begin{subfigure}{0.45\textwidth}
        \includegraphics[width=\textwidth]{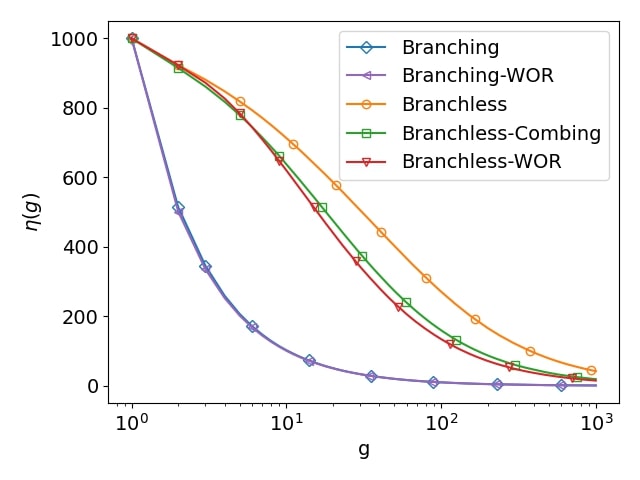}
        \caption{System 1}
        \label{fig:hom_50_K}
    \end{subfigure}
    \begin{subfigure}{0.45\textwidth}
        \includegraphics[width=\textwidth]{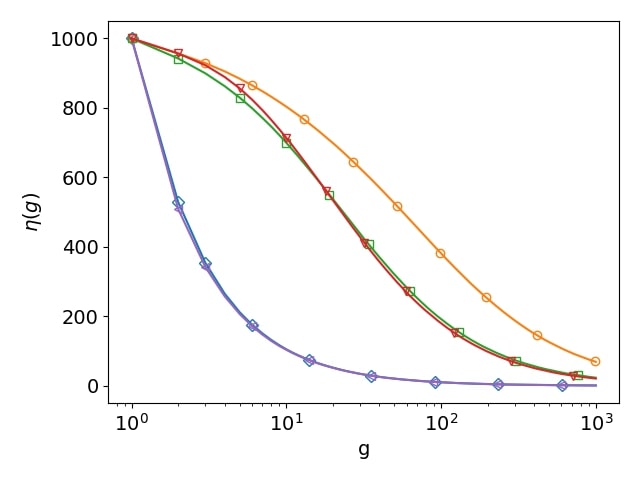}
        \caption{System 2}
        \label{fig:hom_100_K}
    \end{subfigure}
    \begin{subfigure}{0.45\textwidth}
        \includegraphics[width=\textwidth]{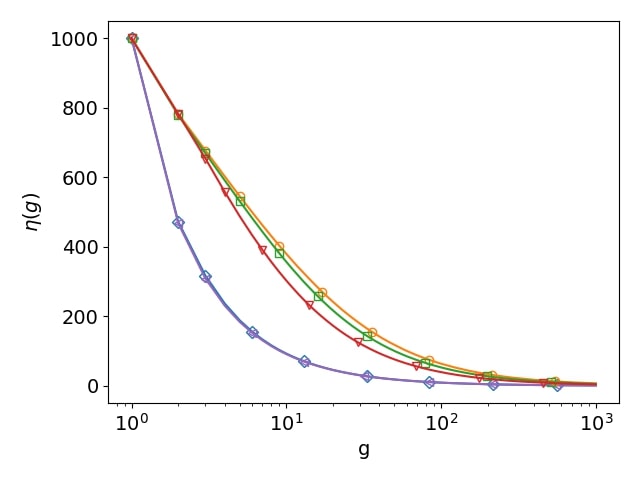}
        \caption{System 3}
        \label{fig:het_42_K}
    \end{subfigure}
    \begin{subfigure}{0.45\textwidth}
        \includegraphics[width=\textwidth]{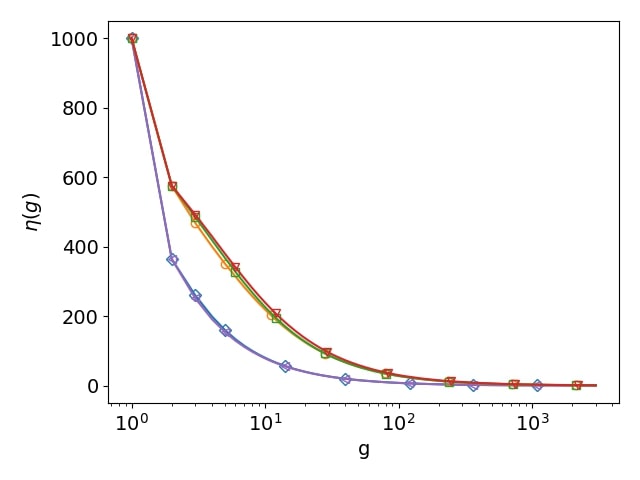}
        \caption{System 4}
        \label{fig:het_60_K}
    \end{subfigure}
    \caption{Average number of surviving families as a function of the generations, computed for $N=10^3$ and $B=10^2$, for different systems, with $10^3$ independent replicas.}
    \label{fig:pi_families}
\end{figure}

\begin{figure}[t]
    \centering
    \begin{subfigure}{0.45\textwidth}
        \includegraphics[width=\textwidth]{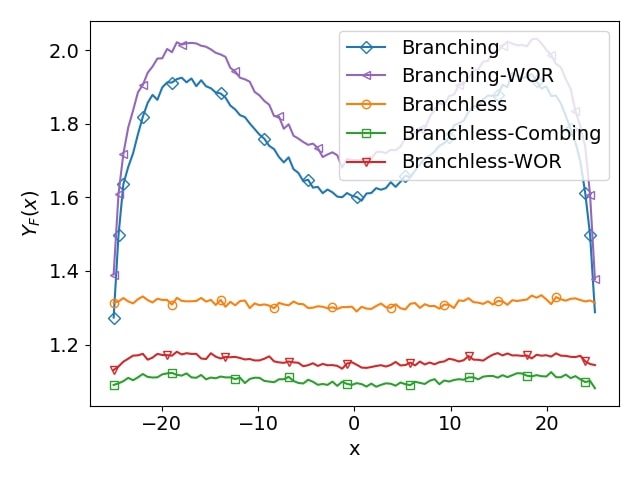}
        \caption{$Y_F$ in power iteration for system 1 with $10^5$ independent replicas.}
        \label{fig:hom_50_YF}
    \end{subfigure}
    \begin{subfigure}{0.45\textwidth}
        \includegraphics[width=\textwidth]{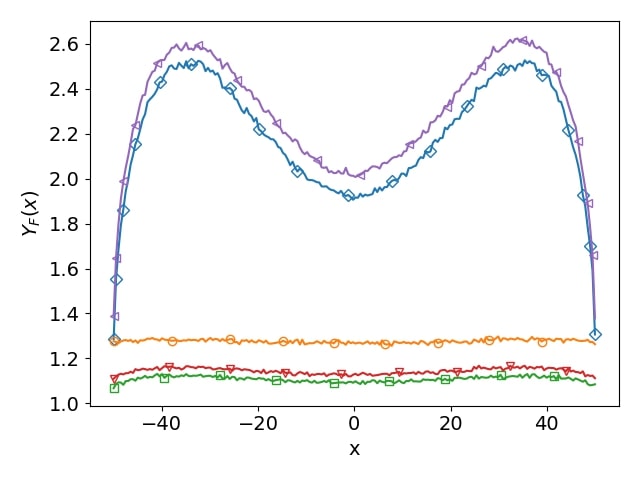}
        \caption{$Y_F$ in power iteration for system 2 with $10^5$ independent replicas.}
        \label{fig:hom_100_YF}
    \end{subfigure}
    \begin{subfigure}{0.45\textwidth}
        \includegraphics[width=\textwidth]{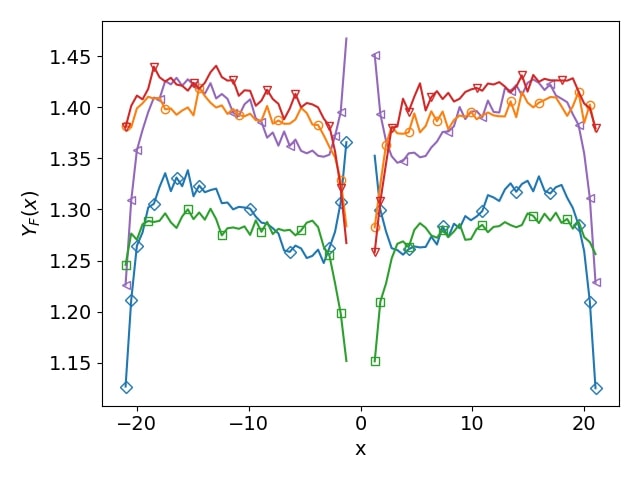}
        \caption{$Y_F$ in power iteration for system 3 with $10^4$ independent replicas.}
        \label{fig:het_42_YF}
    \end{subfigure}
    \begin{subfigure}{0.45\textwidth}
        \includegraphics[width=\textwidth]{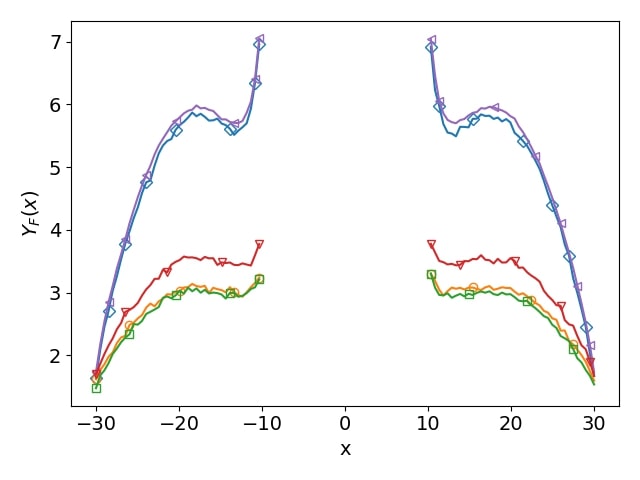}
        \caption{$Y_F$ in power iteration for system 4 with $10^4$ independent replicas.}
        \label{fig:het_60_YF}
    \end{subfigure}
    \caption{Equilibrium Feynman moment for the fission source computed for $N=1000$ and $\delta=0.5$ for different systems.}
    \label{fig:pi_Yfiss}
\end{figure}

Careful inspection of Fig.~\ref{fig:pi_entropy} will reveal that the difference between the asymptotic entropy of the branching algorithm and of the branchless algorithms increases with the dominance ratio. Conversely, if the dominance ratio is small enough, the entropy for branching collisions becomes slightly higher than for branchless collisions, as illustrated by Fig.~\ref{fig:het_42_S} for system 3. In this case, correlated branches explore the whole viable phase space, and the effects of spatial clustering will be quenched. Said differently, neutrons produce more fission neutrons (i.e.\ correlated branches) when using branching collisions than branchless collisions, meaning that more random walkers are available to explore the phase space. Given that spatial correlations are weak, the increasing number of events leads to better estimates of scores. This behavior is not specific to heterogeneous systems (such as configuration 3): we have found similar results for a homogeneous configuration with a dominance ratio close to that of system 3. Note however that, in systems with small dominance ratios, the difference among the asymptotic entropy values for the five algorithms is rather small, translating the fact that systems with small dominance ratio are anyway only weakly affected by spatial correlations and, thus, by clustering.

The findings in Fig.~\ref{fig:pi_pair_distance} for the pair distance function are consistent with those stemming from the entropy tally, with one significant exception: contrary to entropy, population control seems to have a weak influence on the pair correlation function. Moreover, Fig.~\ref{fig:het_42_R} shows that the pair distance for the branching algorithm in system 3 is the lowest amongst the five algorithms considered here (the differences are nonetheless small). This is in contrast with the intuition that lower entropy values are the signature of enhanced spatial clustering. We showed above that, for system 3, the highest entropy (and thus, the situation with the smallest impact of neutron clustering) was also achieved for the branching algorithm. Consequently, we would have expected that the largest pair distance function would be attained for the branching algorithm. Nonetheless, system 3 will not exhibit significant clustering for any reasonable number of particles, mostly because of its low dominance ratio. 

For systems exhibiting strong heterogeneity, such as system 4, the use of the average pair distance tally to assess the impact of spatial correlations in the fission source becomes questionable, since the moderator region does not contain fission neutrons: if the pair distance is comparable to the linear size of the moderator region, then it is unclear whether the pair distance simply encodes the distance between the two fissile regions.

Figure~\ref{fig:pi_families} shows the average number of surviving families as a function of generations. At the fixation generation, all neutrons in the population are correlated: a larger fixation generation should imply weaker correlations. Figures~\ref{fig:hom_50_K} and \ref{fig:hom_100_K} contradict this statement: in homogeneous media, branchless collisions without population control lead to the largest fixation generation, whereas the highest entropy and pair distance function (hence the weakest spatial correlations) are reached for branchless collisions with combing. Similarly, Fig.~\ref{fig:het_42_K} shows that branching collisions lead to the smallest fixation generation, although the associated entropy is the highest amongst the algorithms considered here. This surprising finding is coherent with previous work: the statistics of families are unsuitable to detect correlations in systems of finite size~\cite{bonnet_fixation_2023}. In infinite media, the dynamics of spatial correlations is driven by the statistics of families, due to the lack of a spatial scale. On the contrary, in systems of finite size, the emergence of spatial clustering depends on the competition between the number of generations required to reach fixation and the number of generations required to reach mixing, i.e.\ to explore the whole space~\cite{mulatier_critical_2015,zoia_neutron_2017,bonnet_space_2022}. In this context, the statistics of families alone becomes insufficient to provide the full picture.

\subsection{Analysis of the fission source: Feynman moment}
\label{subsubsection:fission_feynman}

In view of the shortcomings of global tallies, we move now to the analysis of local tallies: the Feynman moment and the normalized variance. These tallies will be applied to the fission source and to the collision counts. We will investigate both the spatial shape of the local tallies and their dependence on the mesh cell size $\delta$. Indeed, coarse-graining over a length $\delta$ influences the impact of correlations on the sought observables. 

The Feynman moment of the fission source describes the correlations between fission neutrons. Figure~\ref{fig:pi_Yfiss} shows the ensemble Feynman moments for the fission source after the inactive generations, for each configuration and algorithm considered. Let us initially focus on the homogeneous cases, i.e.\ systems 1 and 2. In the ideal case where all particles are independent, we expect fission neutrons to be Poisson distributed, which is equivalent to a spatially flat Feynman moment $Y \simeq 1$ everywhere. Figures~\ref{fig:hom_50_YF} and \ref{fig:hom_100_YF} show that the Feynman moment for the branching algorithm in a homogeneous configuration has a non-trivial shape, indicating that deviations from the ideal behavior do not develop uniformly. The same figures illustrate that simulations using branchless collisions without additional population control introduce smaller and more evenly distributed fluctuations in the distribution of fission sites, as witnessed by the lower and spatially flatter Feynman moment. The discrepancy in the Feynman moment for branchless and branching collisions increases with increasing dominance ratio: for branching collisions $Y$ increases with increasing dominance ratio, whereas it is fairly constant for branchless collisions when increasing the dominance ratio. Introducing combing or sampling WOR in branchless algorithms lowers the Feynman moments and introduces a slight spatial shape, which suggests that population control algorithms amplify the effects of leakage boundary conditions on spatial correlations. This is reminiscent of what happens with reflecting boundary conditions~\cite{sutton_application_2017}. On the contrary, introducing sampling WOR in branching collisions appears to increase the Feynman moments. This is in apparent contradiction with Sutton's conclusions \cite{sutton_toward_2022}. However, we have verified that the discrepancy with the works in Ref.~\citenum{sutton_toward_2022} appears to be related to going from analog to non-analog collisions, and from a mono-kinetic to a multi-group framework. In fact, in the systems investigated in Ref.~\citenum{sutton_toward_2022}, sampling WOR introduces no weight fluctuations when sampling fission neutrons: all fission sites are equivalent (same weight, same energy), which does not hold true in the benchmark problems chosen here. This raises questions about alternative extensions of sampling WOR that may be better suited to the multi-group framework and to non-analog algorithms.

\begin{figure}[t]
    \centering
    \begin{subfigure}{0.49\textwidth}
        \includegraphics[width=\textwidth]{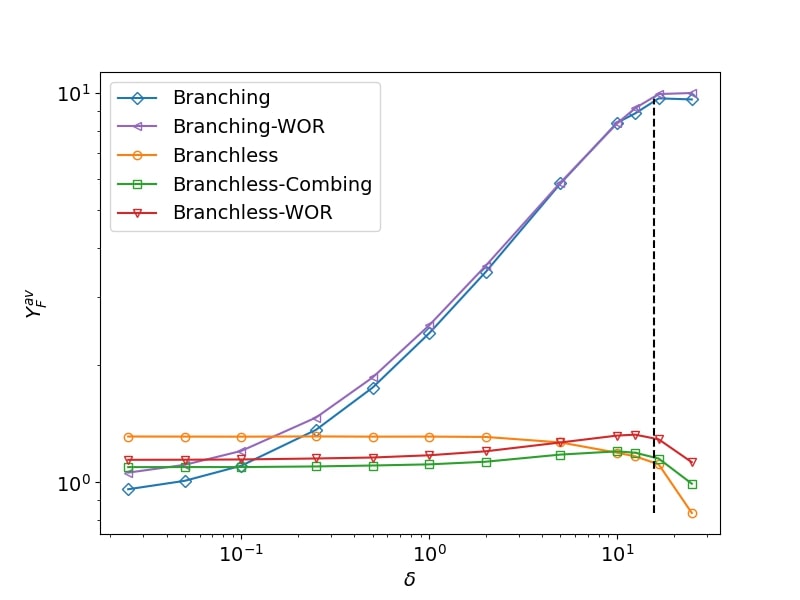}
        \caption{System 1 with $10^5$ realizations.}
        \label{fig:Yfiss_wrt_B_hom_50}
    \end{subfigure}
    \begin{subfigure}{0.49\textwidth}
        \includegraphics[width=\textwidth]{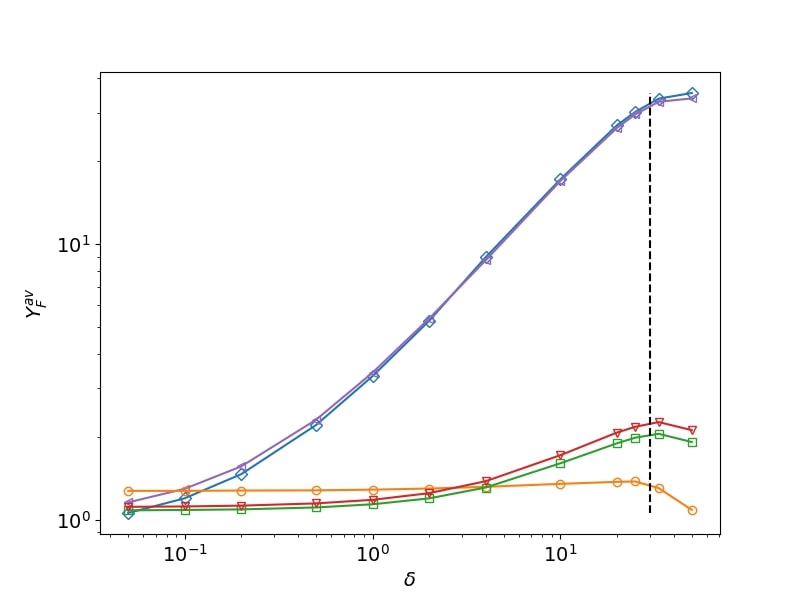}
        \caption{System 2 with $10^5$ realizations.}
        \label{fig:Yfiss_wrt_B_hom_100}
    \end{subfigure}
    \begin{subfigure}{0.49\textwidth}
        \includegraphics[width=\textwidth]{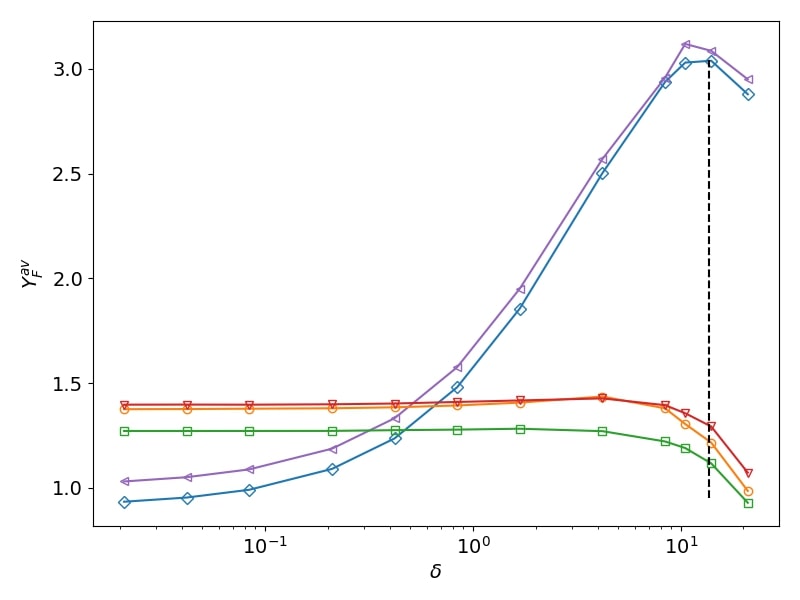}
        \caption{System 3 with $10^5$ realizations.}
        \label{fig:Yfiss_wrt_B_het_42}
    \end{subfigure}
    \begin{subfigure}{0.49\textwidth}
        \includegraphics[width=\textwidth]{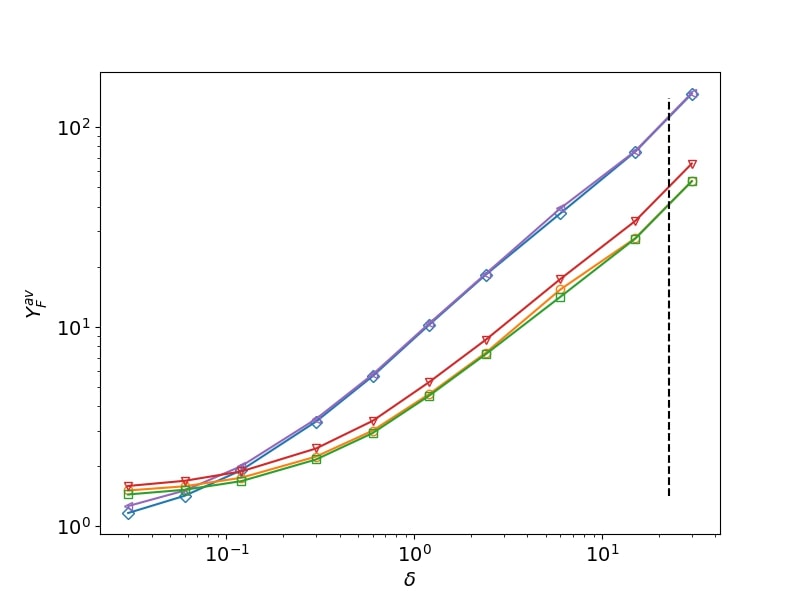}
        \caption{System 4 with $10^4$ realizations.}
        \label{fig:Yfiss_wrt_B_het_60}
    \end{subfigure}
    \caption{Space-averaged Feynman moment of the fission source, as a function of the cell size $\delta$, for $N=10^3$. Dashed black line: square root of $\langle r^2 \rangle_\infty$ for branching algorithm. Given that the $\langle r^2 \rangle_\infty$ are relatively close to each other for all five algorithms, we only show the value for the branching algorithm.}
    \label{fig:space_YFiss}
\end{figure}

Insight can be gained by examining how $Y$ scales with respect to the cell size $\delta$. We define $l_\text{clust}$ as the typical size of the largest cluster in the system. In systems that are homogeneous or where the effects of heterogeneity on particle transport are mild, we have that $l_\text{clust} \sim \sqrt{\langle r^2 \rangle_\infty}$; this assumption fails for systems where the effects of heterogeneity are more pronounced, such as configuration 4. We thus have the following picture: if $\delta \simeq l_\text{clust}$, the observables will be strongly affected by correlations; on the contrary, if $\delta \ll l_\text{clust}$, the probability that two particles belonging to the same cluster fall into the same cell becomes small, and the effects of the correlations will be milder. Figures~\ref{fig:Yfiss_wrt_B_hom_50} and \ref{fig:Yfiss_wrt_B_hom_100} show the Feynman moment for the spatially-averaged fission source defined by
 \begin{equation}
     Y^{av}_F = \frac{1}{B'}\sum_{i=1}^{B'} Y_F^i,
     \label{eq:feynman_averaged}
 \end{equation}
where $B'$ is the number of spatial cells with fissile material. As we consider homogeneous systems, $l_\text{clust} \sim \sqrt{\langle r^2 \rangle_\infty}$. The Feynman moment for branching collisions has a power law increase for $\delta \lesssim \sqrt{\langle r^2 \rangle_\infty}$ and saturates when $\delta \sim \sqrt{\langle r^2 \rangle_\infty}$. This translates the fact that there exists clusters at all spatial scales up to $\sqrt{\langle r^2 \rangle_\infty}$. In addition, we observe that adding sampling WOR does not significantly alter Feynman moments for $\delta \sim \sqrt{\langle r^2 \rangle_\infty}$, and leads to slightly higher Feynman moments when $\delta \ll \sqrt{\langle r^2 \rangle_\infty}$. This suggests that sampling WOR actually favors short-distance correlations. More strikingly, the Feynman moment for branchless collisions without population control is constant when $\delta \lesssim \sqrt{\langle r^2 \rangle_\infty}$ and decreases when $\delta \sim \sqrt{\langle r^2 \rangle_\infty}$, the latter being related to the counts following a binomial law instead of a Poisson law when the number of bins is small, as mentioned earlier. In this case, clustering is completely mitigated by the use of branchless collisions: if fission neutrons are Poisson distributed, the variance and the average of the distribution of fission neutrons scale similarly with respect to $\delta$ and the Feynman moment becomes constant. When adding combing or sampling WOR, the Feynman moment stays constant for $\delta \ll \sqrt{\langle r^2 \rangle_\infty}$ and start increasing when approaching $\delta \sim \sqrt{\langle r^2 \rangle_\infty}$, whereupon it saturates.

The shape of the curves in Fig.~\ref{fig:space_YFiss} can be given an interesting interpretation. When the average Feynman moment is constant for small values of $\delta$, as for all branchless algorithms, it translates the fact that clusters have some minimum spatial scale; at scales finer than the minimum cluster size, clusters are essentially infinitely large, and the neutrons occupation statistics behaves as if the particles were independent. At the opposite extreme, the Feynman moment saturates at scales of the order of $\delta \sim \sqrt{\langle r^2 \rangle_\infty}$, which (at least in homogeneous systems) can be interpreted as the typical size of the largest clusters. Between the minimum and maximum scales, the Feynman moment follows a power law, which is the expected behavior for self-organized critical systems, as recently suggested by Dechenaux et al.\ on the basis of the application of techniques borrowed from the renormalization group in particle physics~\cite{dechenaux_percolation_2022}. The power-law exponents range between $0.5$ and $1$ for systems 1 and 2. The occurrence of power laws in the scale dependence of Feynman spectra is suggestive, but more careful analysis is required to confirm the existence of a deeper connection. Finally, at scales coarser than the typical cluster size, the occupation statistics of the space bins behaves as if the system were populated by independent macro-particles. In the limit $\delta\to L$, the Feynman moment decreases, as the counts follow binomial laws, and converges to zero because of the constraint on weight conservation of the fission source.

We move now to the case of heterogeneous systems 3 and 4. Figure~\ref{fig:het_42_YF} shows the Feynman moments for system 3, which has a small dominance ratio. For this case, all algorithms behave similarly. The main difference lies at the fuel-moderator interfaces. Here, branchless collisions lead to a dip in the Feynman moment near the material boundary, and a peak for branching collisions. Figure~\ref{fig:het_60_YF} shows that for case 4, which has a larger dominance ratio, the Feynman moment is significantly smaller when using branchless collisions than when using branching collisions. The spatial shape of the Feynman moments is similar in all heterogeneous cases, suggesting that spatial correlations are always at play, although they may be weaker for branchless collisions. Additionally, the Feynman moment using sampling WOR is higher than the other variants in both heterogeneous systems, casting some doubt on the usefulness of sampling WOR in heterogeneous media.

Again, interesting conclusions can be drawn from the dependence of the Feynman moment on $\delta$. Figure~\ref{fig:Yfiss_wrt_B_het_42} shows that the \textit{spatially-averaged} Feynman moment for all branchless variants in the heterogeneous configuration with a small dominance ratio is roughly constant for $\delta \lesssim \sqrt{\langle r^2 \rangle_\infty}$, and decreases for $\delta \gtrsim \sqrt{\langle r^2 \rangle_\infty}$. This suggests the absence of strong spatial correlations when using any of the five algorithms; it also illustrates that the pair distance function is still a good approximation of the typical cluster size for systems with mild heterogeneity. On the other hand, Fig.~\ref{fig:Yfiss_wrt_B_het_60} shows that $Y_F^{av} \propto \delta^{0.82}$ for $\delta \lesssim \sqrt{\langle r^2 \rangle_\infty}$ for all algorithms in the heterogeneous configuration with a large dominance ratio. Moreover, the power law scaling of the space-averaged Feynman moment as a function of $\delta$ is the same for each algorithm. By analogy with the three other cases, we expect the space-averaged Feynman moment in system 4 to reach a maximum for $\delta \gtrsim l_\text{clust}$; a direct confirmation is not possible, because $\sqrt{\langle r^2 \rangle_\infty}$ is essentially driven by the size of the moderator region, which means that it is not a good approximation of the typical cluster size, which is unknown.

So far, we have investigated ensemble Feynman moments, which are not directly accessible by power iteration (unless one accepts to sample independent replicas). One may wonder how the ergodic Feynman moment relates to the ensemble Feynman moment. We consider then the dependence of the Feynman moment on the number of active generations $G$: based on Refs.~\citenum{miao_predicting_2018} and \citenum{mickus_does_2021}, we expect that, when the number of active generations is large enough, the apparent variance converges towards the real variance. Therefore, we also expect the ergodic Feynman moments to converge towards the ensemble Feynman moments. In what follows, we want to characterize the relationship between the value of the dominance ratio and the ability of our algorithms to mitigate the bias on the variance induced by the use of ergodic averages instead of ensemble averages. We stress this bias is \textit{not} related to discarding an insufficient number of inactive generations before tallying the observables.

\begin{figure}[t]
    \centering
    \begin{subfigure}{0.45\textwidth}
        \includegraphics[width=\textwidth]{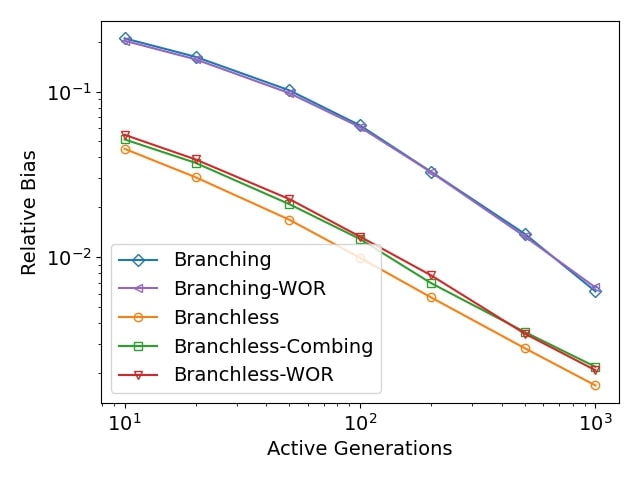}
        \caption{System 1.}
        \label{fig:Yfiss_wrt_G_hom_50}
    \end{subfigure}
    \begin{subfigure}{0.45\textwidth}
        \includegraphics[width=\textwidth]{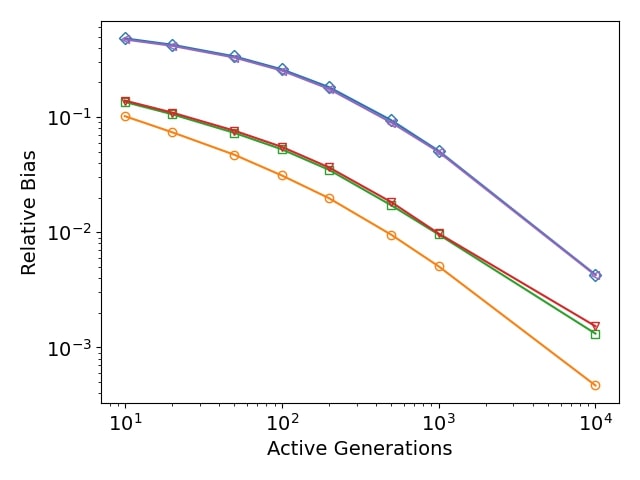}
        \caption{System 2.}
        \label{fig:Yfiss_wrt_G_hom_100}
    \end{subfigure}
    \begin{subfigure}{0.45\textwidth}
        \includegraphics[width=\textwidth]{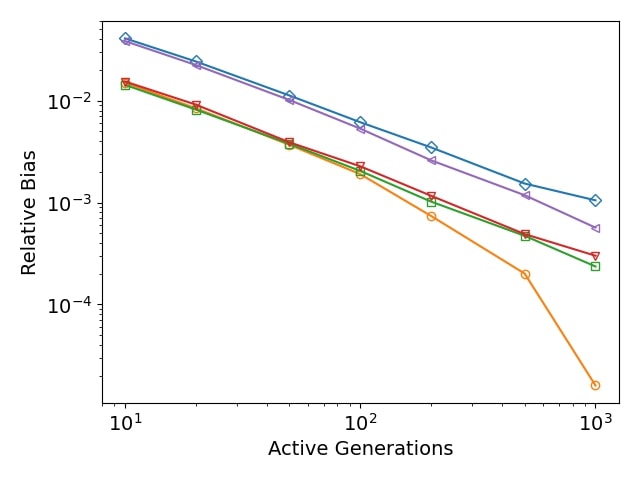}
        \caption{System 3.}
        \label{fig:Yfiss_wrt_G_het_42}
    \end{subfigure}
    \begin{subfigure}{0.45\textwidth}
        \includegraphics[width=\textwidth]{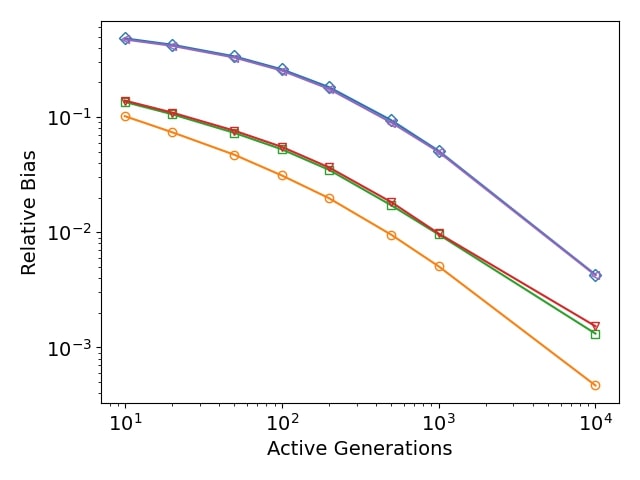}
        \caption{System 4.}
        \label{fig:Yfiss_wrt_G_het_60}
    \end{subfigure}
    \caption{Convergence of the space averaged $Y_F^{G,av}$ with respect to the number of active generations $G$ in terms of the relative bias $|Y^{av}_F - Y^{G,av}_F|/Y^{av}_F$. The population size is $N=10^3$ and $B=10^2$. Same legend as Fig.~\ref{fig:pi_Yfiss}.}
    \label{fig:Yfiss_wrt_G}
\end{figure}

Figures~\ref{fig:Yfiss_wrt_G_hom_50}-\ref{fig:Yfiss_wrt_G_het_60} show the convergence of the spatially-averaged Feynman moments $Y_F^{G,av}$ obtained by ergodic average over $G$ generations towards the ensemble spatially-averaged Feynman moments $Y_F^{av}$. For this purpose, we compute the relative bias defined by $|Y^{av}_F - Y^{G,av}_F|/Y^{av}_F$, as a function of the number of active generations. In all cases, it is clear that the underestimation of errors due to ergodic averages is significantly smaller when using branchless collisions than when using branching collisions, by about a factor $5$ in homogeneous systems and up to a factor $3$ in heterogeneous systems. Additionally, the slope of the relative bias as a function of active generations only weakly depends on the details of the sampling algorithm and population control, and is essentially driven by the value of the dominance ratio. However, the `initial offset' depends on sampling and population control. This is reminiscent of the fact that, in power iteration, the speed of convergence of $k_0$ only depends on the dominance ratio, but the initial value of $k_0$ (and so the number of inactive generations to discard) depends on the initial condition. In addition, the effect of population control algorithms appears to be milder compared to the effect of collision biasing.

\begin{figure}[t]
    \centering
    \begin{subfigure}{0.45\textwidth}
        \includegraphics[width=\textwidth]{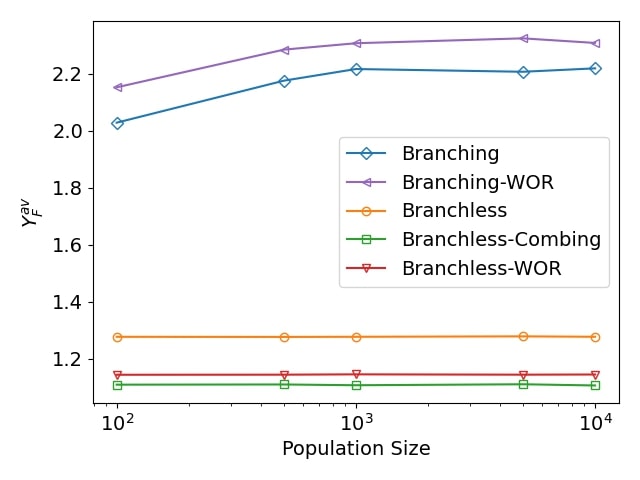}
        \caption{For homogeneous configuration with large dominance ratio (system 2)}
        \label{fig:Yfiss_wrt_N_hom_100}
    \end{subfigure}
    \begin{subfigure}{0.45\textwidth}
        \includegraphics[width=\textwidth]{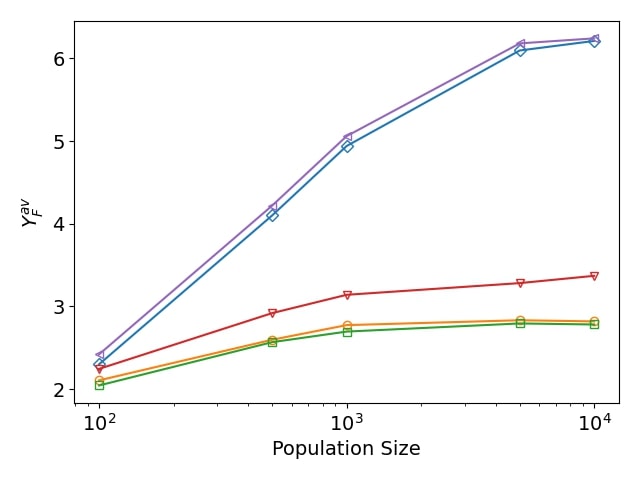}
        \caption{For heterogeneous configuration with large dominance ratio (system 4)}
        \label{fig:Yfiss_wrt_N_bl}
    \end{subfigure}
    \caption{Space-averaged ensemble Feynman moments of the fission source $Y_F^{av}$ as a function of the population size $N$, with bin size $\delta = 0.5~\unit{cm}$.}
\end{figure}

We conclude our analysis by considering the effects of the neutron population size. Previous investigations have shown that the impact of spatial clustering becomes milder as the number of particles per generation increases~\cite{mulatier_critical_2015}. Figures~\ref{fig:Yfiss_wrt_N_hom_100}-\ref{fig:Yfiss_wrt_N_bl} show the space-averaged ensemble Feynman moments for all five algorithms in system 2 (homogeneous case with higher dominance ratio), as a function of the population size. We observe that the Feynman moment for the branching algorithm converges towards an asymptotic value for large $N$. This scaling is consistent with similar findings in the context of time-dependent Monte Carlo simulations, for the case of spatially homogeneous branching-diffusion processes under population control~\cite{zoia_neutron_2017}. Additionally, we verified that the shape of the Feynman moment also converges to an asymptotic shape for large $N$ (not showed here for conciseness). We further observe that, using branchless collisions in systems 1-3, Feynman moments appear to be independent of $N$ (only shown for system 2 in Fig.~\ref{fig:Yfiss_wrt_N_hom_100}). This behavior is presumably due to the absence of significant spatial correlations with these algorithms. Conversely, in system 4, where strong spatial correlations exist for all variants of branchless algorithms, the Feynman moments depend on $N$ and converge towards an asymptotic value for large $N$. It should be noted that, even then, the Feynman moment of branchless variants saturates at a smaller value of $N$ than for the branching algorithm.

\begin{figure}[t]
    \centering
    \begin{subfigure}{0.45\textwidth}
        \includegraphics[width=\textwidth]{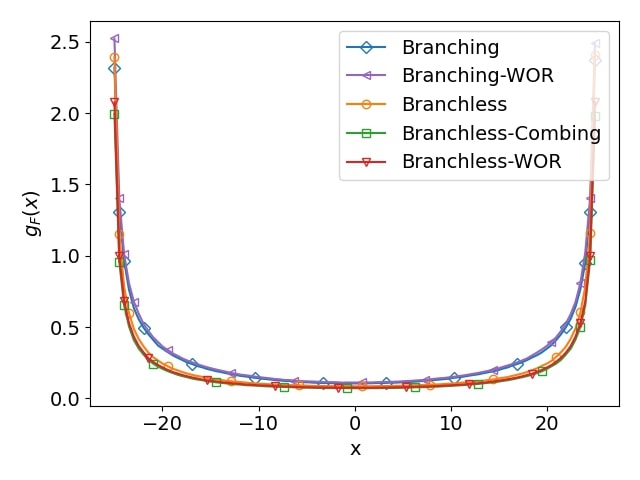}
        \caption{System 1, with $10^5$ independent replicas.}
        \label{fig:hom_50_gfiss}
    \end{subfigure}
    \begin{subfigure}{0.45\textwidth}
        \includegraphics[width=\textwidth]{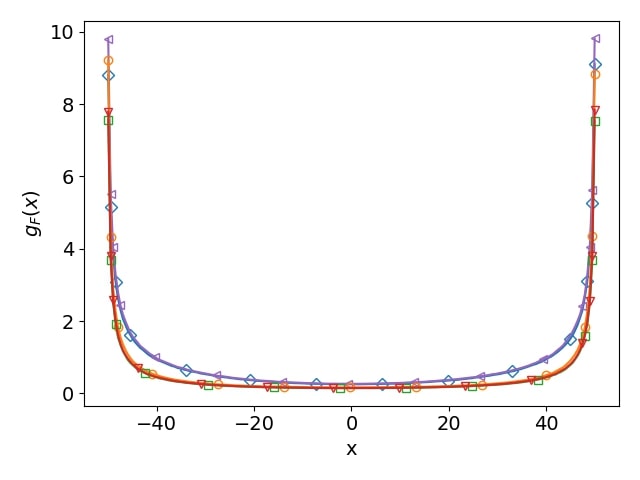}
        \caption{System 2, with $10^5$ independent replicas.}
        \label{fig:hom_100_gfiss}
    \end{subfigure}
    \begin{subfigure}{0.45\textwidth}
        \includegraphics[width=\textwidth]{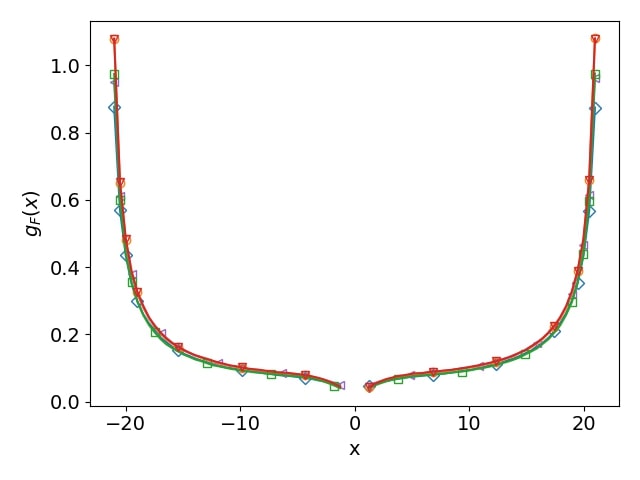}
        \caption{System 3, with $10^5$ independent replicas}
        \label{fig:het_42_gfiss}
    \end{subfigure}
    \begin{subfigure}{0.45\textwidth}
        \includegraphics[width=\textwidth]{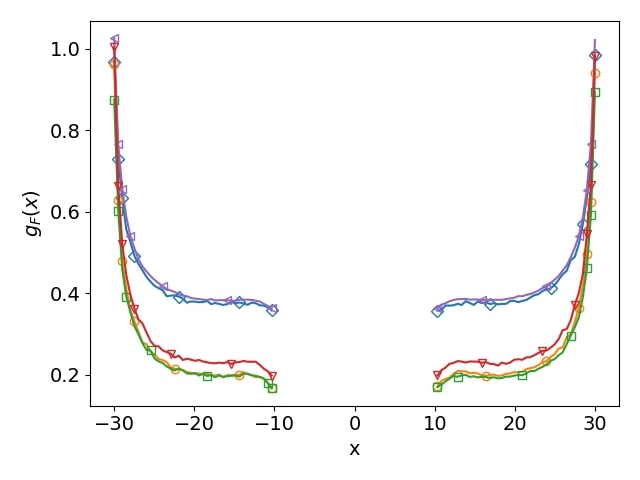}
        \caption{System 4, with $10^4$ independent replicas.}
        \label{fig:het_60_gfiss}
    \end{subfigure}
    \caption{Equilibrium ensemble normalized variances for the fission source computed over $10^5$ independent replicas. $N=1000$ and $B=100$ for different systems.}
    \label{fig:pi_gfiss}
\end{figure}

\subsection{Analysis of the fission source: normalized variance}
\label{subsubsection:fission_variance}

Additional insight can be obtained by assessing the behavior of the \textit{normalized variance} $g_F$ of the fission source. As discussed above, $g\sim0$ is a reasonable definition of `weak correlations', similarly as for $Y\sim1$. However, in the following we will show that the two tallies are not equivalent, and convey different pieces of information. The amplitude of fission-induced spatial fluctuations depends on the population size, and we will illustrate that this is captured by normalized variances.

All cases in Fig.~\ref{fig:pi_gfiss} display a significant increase of $g_F$ close to the leakage boundary conditions, even when using branchless collisions. This peak is explained by inspecting Fig.~\ref{fig:pi_Yfiss}: although Feynman moments are (almost) flat for branchless collisions, the average of the fission source is close to zero near leakage boundary conditions. Since $g_F = Y_F / \mathbb{E}[F]$, this explains the shape of $g_F$.

Let us analyze the homogeneous systems first. In Figs.~\ref{fig:hom_50_gfiss}-\ref{fig:hom_100_gfiss} we observe that the normalized variances of all branchless algorithms are rather similar, and close to zero in the bulk of the geometry. On the other hand, the normalized variance using the branching algorithm can be close to $1$ even far from the boundaries, which suggests that large fluctuations will occur. We observe that the shape of the normalized variance for all algorithms is almost independent of the dominance ratio. In the case of the heterogeneous media, Fig.~\ref{fig:het_42_gfiss} for case 3 shows that all the algorithms behave similarly in terms of normalized variances. Furthermore, Fig.~\ref{fig:het_60_gfiss} for case 4 confirms the hierarchy between the algorithms, and shows that, even in the bulk of the system, typical fluctuations are larger for the branching algorithm than for the branchless algorithms. Similarly to the homogeneous cases, the normalized variance is only weakly dependent on the dominance ratio using branchless algorithms, except for sampling WOR. The reason why sampling WOR behaves worse in heterogeneous media is still unclear and would require further investigation.

\begin{figure}[t]
    \centering
    \begin{subfigure}{0.45\textwidth}
        \includegraphics[width=\textwidth]{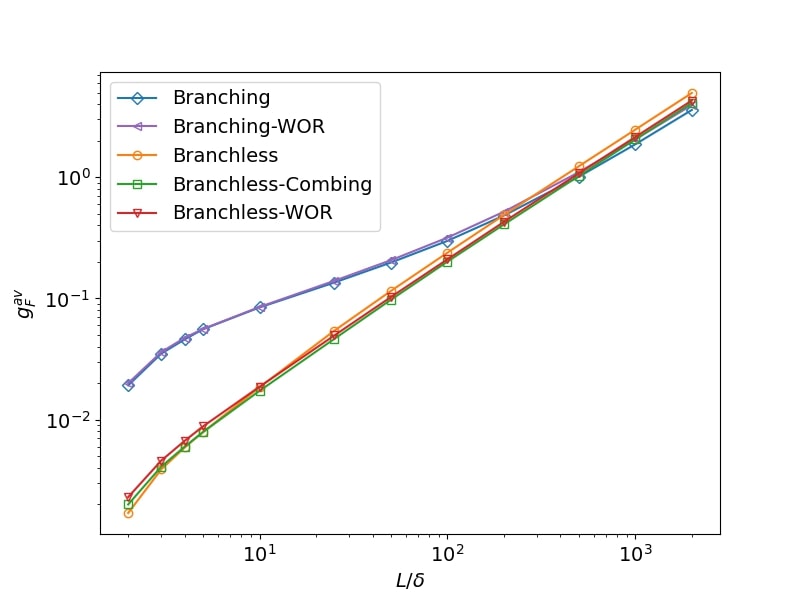}
        \caption{System 1.}
        \label{fig:scaling_B_gfiss_hom_50}
    \end{subfigure}
    \begin{subfigure}{0.45\textwidth}
        \includegraphics[width=\textwidth]{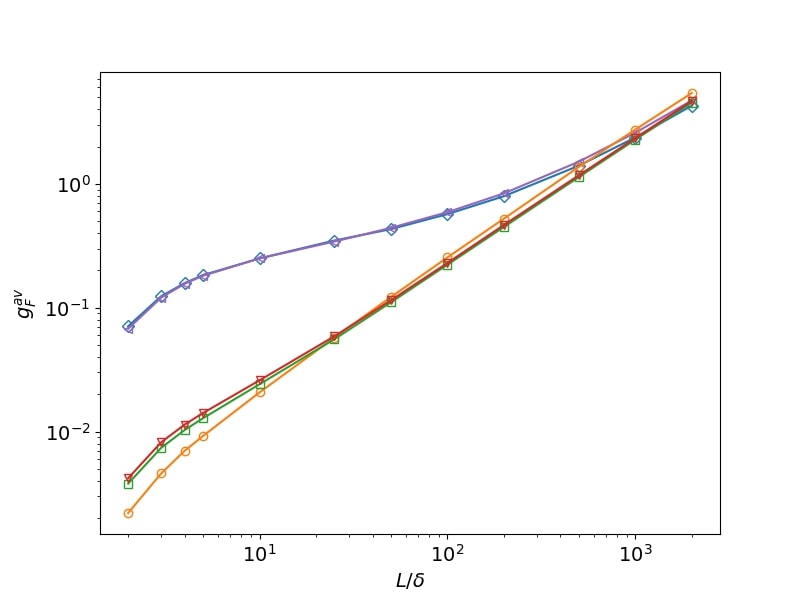}
        \caption{System 2.}
        \label{fig:scaling_B_gfiss_hom_100}
    \end{subfigure}
    \begin{subfigure}{0.45\textwidth}
        \includegraphics[width=\textwidth]{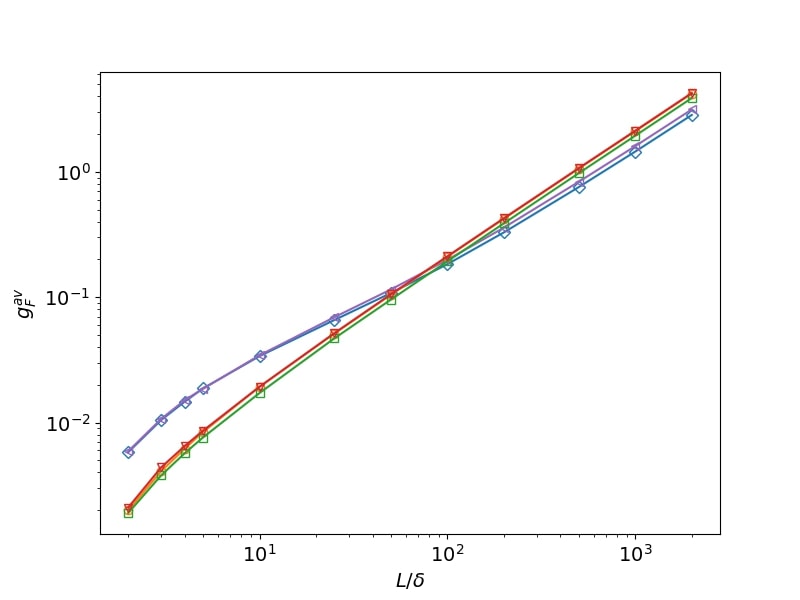}
        \caption{System 3.}
        \label{fig:scaling_B_gfiss_het_42}
    \end{subfigure}
    \begin{subfigure}{0.45\textwidth}
        \includegraphics[width=\textwidth]{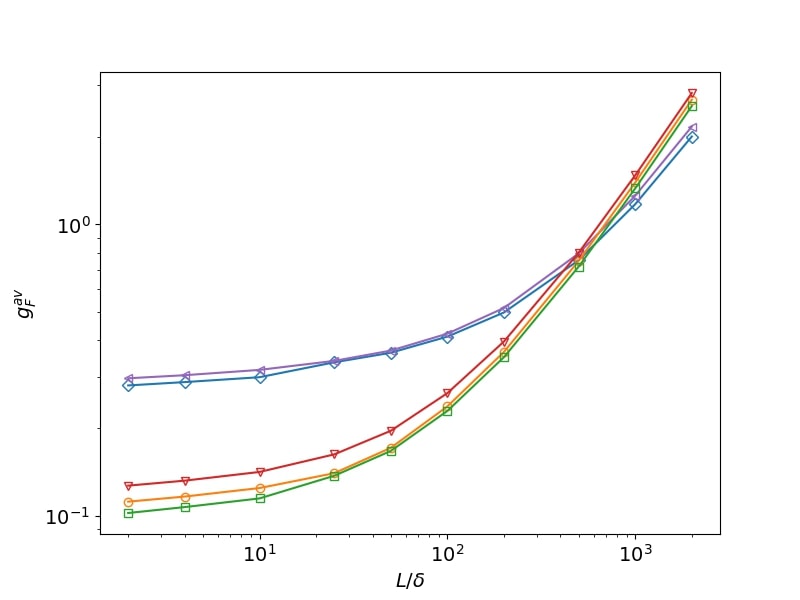}
        \caption{System 4.}
        \label{fig:scaling_B_gfiss_het_60}
    \end{subfigure}
    \caption{Scaling of the space-averaged normalized variance $g_F^{av}$ as a function of the number of cells $L/\delta = B$, for all 4 systems and algorithms, $N=10^3$, and with $10^4$ independent replicas.}
    \label{fig:B_dependence_gfiss}
\end{figure}

Next, we examine the dependence of normalized variances on the population size $N$ and the cell size $\delta$. For the sake of conciseness, we will not consider the dependence on $G$, the number of active generations, because these effects are very similar to those occurring for Feynman moments of Sec.~\ref{subsubsection:fission_feynman}. As for the dependence on $\delta$, previous findings suggest that $g_F \sim 1/\delta$ for small values of $\delta$ \cite{zoia_neutron_2017}. Figure~\ref{fig:B_dependence_gfiss} illustrates this scaling. We consider
\begin{equation}
    g^{av}_F(B) = \frac{1}{B'}\sum_{i=1}^{B'} g_F^i,
\end{equation}
where $B'$ is the number of cells with fissile material. We note that $g^{av}_F \propto B$ for all algorithms, when $L/\delta  = B \gg 1$. For large cell sizes, the normalized variance is smaller for branchless collisions than for branching collisions. The converse is true for small cell sizes. This can be explained by the fact that spatial correlations are weakened by decreasing the cell size.
Branching sampling typically results in more fission neutrons than branchless sampling, because each collision site is a potential fission site in branching sampling. Due to the excess of fission neutrons, when spatial correlations are weak, the branching algorithm introduces less variance in the estimation of the fission source than branchless algorithms. There are essentially two ways to obtain weak spatial correlations: either the dominance ratio is small enough (cf.\ system 3) or the cell size is small enough (as clearly illustrated by Figs.~\ref{fig:scaling_B_gfiss_hom_100} and \ref{fig:scaling_B_gfiss_het_60}). Finally, additional numerical findings (not shown here for the sake of conciseness) suggest that the normalized variances scale as $1/N$ regardless of the algorithm, which is consistent with the literature~\cite{mulatier_critical_2015}.

\subsection{Analysis of the collision counts: Feynman moment}

Most of the tallies of interest for reactor physics (such as reaction rates or energy deposition) are estimated by summing over all the collisions in a generation. In this respect, the Feynman moments of the collision counts can help to interpret the behavior of spatial correlations for such observables.

\begin{figure}[t]
    \centering
    \begin{subfigure}{0.45\textwidth}
        \includegraphics[width=\textwidth]{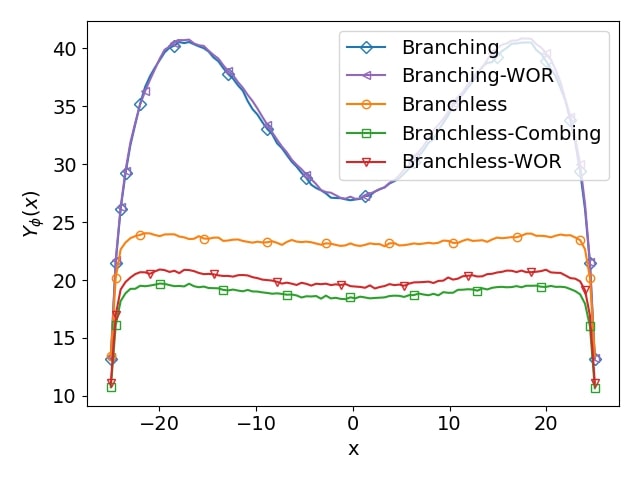}
        \caption{System 1 with $10^5$ independent replicas.}
        \label{fig:hom_50_Ycol}
    \end{subfigure}
    \begin{subfigure}{0.45\textwidth}
        \includegraphics[width=\textwidth]{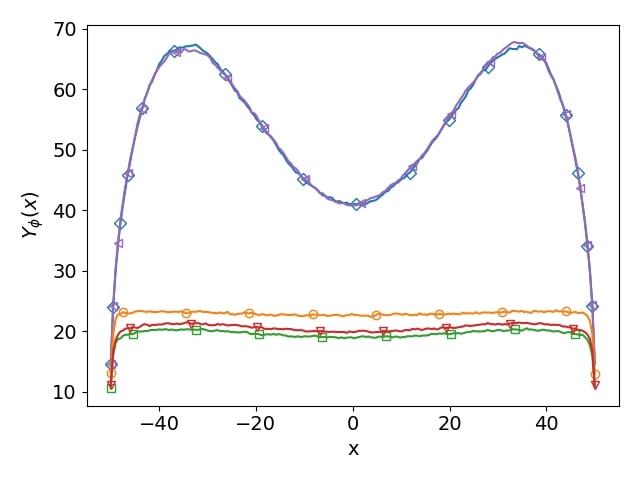}
        \caption{System 2 with $10^5$ independent replicas.}
        \label{fig:hom_100_Ycol}
    \end{subfigure}
    \begin{subfigure}{0.45\textwidth}
        \includegraphics[width=\textwidth]{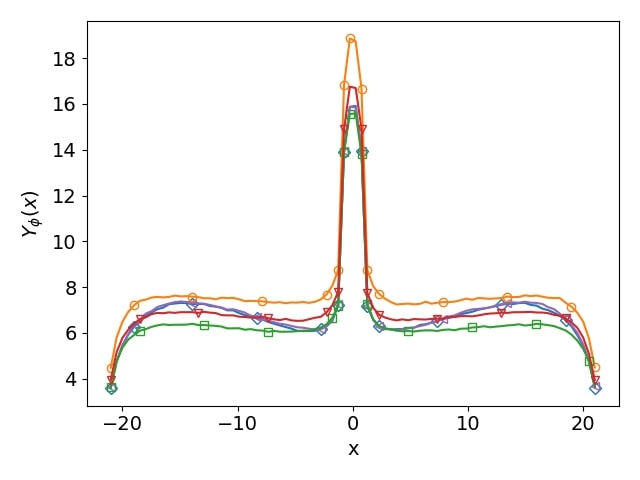}
        \caption{System 3 with $10^4$ independent replicas.}
        \label{fig:het_42_Ycol}
    \end{subfigure}
    \begin{subfigure}{0.45\textwidth}
        \includegraphics[width=\textwidth]{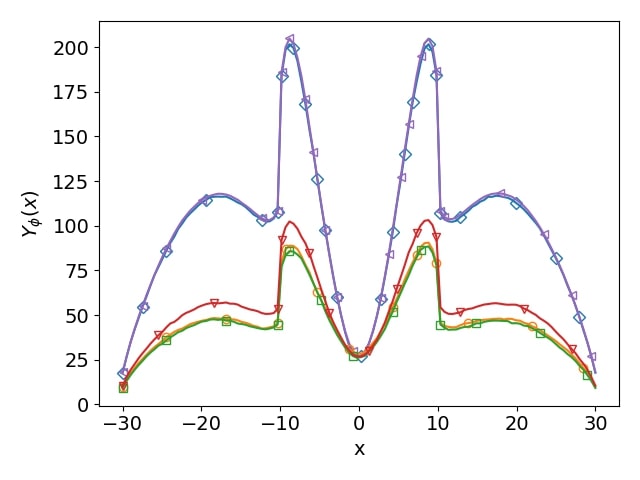}
        \caption{System 4 with $10^4$ independent replicas.}
        \label{fig:het_60_Ycol}
    \end{subfigure}
    \caption{Equilibrium Feynman moments for the collision counts of neutrons computed for $N=1000$ and $B=100$ for different systems.}
    \label{fig:pi_Ycol}
\end{figure}

The Feynman moments of the collision counts, showed in Fig.~\ref{fig:pi_Ycol}, are significantly higher than those of the fission source. Scattering now contributes to the tally and there are significantly more scattering events than fission events; additionally, correlations between collision sites are stronger than correlations between fission sites. Figures~\ref{fig:het_42_Ycol} and \ref{fig:het_60_Ycol} for the heterogeneous cases 3 and 4 indicate strong spatial correlations between collision sites near the fuel-moderator interfaces. Similar peaks in the spatial profile of the correlations were already observed in the fission source, but their amplitude is considerably amplified in the Feynman moments of the collision counts. Far from the fissile regions, all algorithms behave similarly, due to the fact that they all reduce to implicit capture in the moderator region. Note that sampling WOR has no significant effect on the Feynman moments of the collision counts.

\begin{figure}[t]
    \centering
    \begin{subfigure}{0.45\textwidth}
        \includegraphics[width=\textwidth]{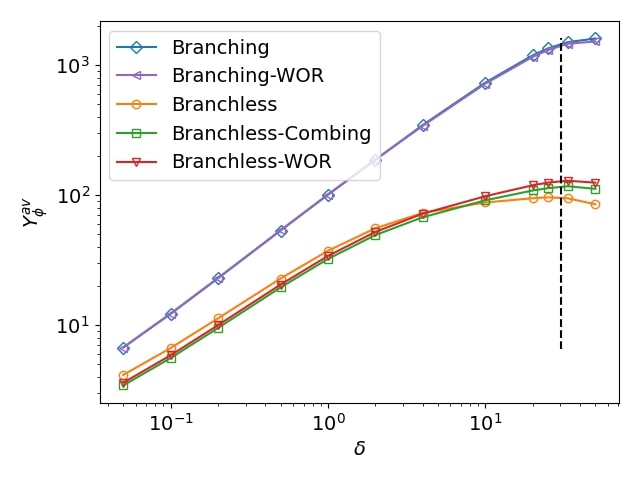}
        \caption{System 2, with $10^4$ replicas.}
        \label{fig:Ycol_wrt_B_hom_100}
    \end{subfigure}
    \begin{subfigure}{0.45\textwidth}
        \includegraphics[width=\textwidth]{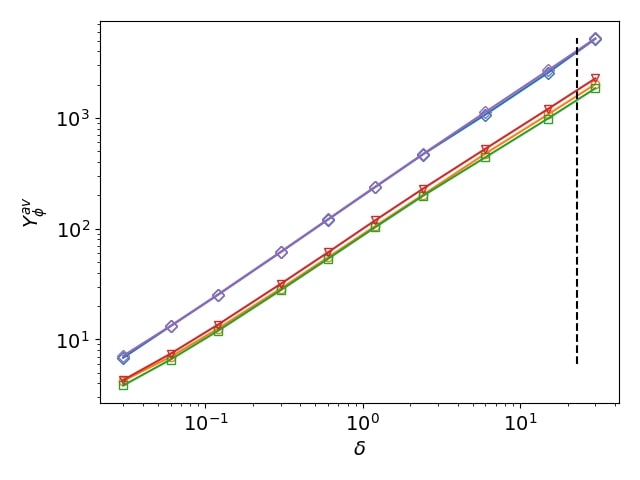}
        \caption{System 4, with $10^3$ replicas.}
        \label{fig:Ycol_wrt_B_het_60}
    \end{subfigure}
    \caption{Space-averaged Feynman moments of the collision counts for all five algorithms, as a function of the cell size $\delta$, for $N=10^3$. Dashed black line: square root of $\langle r^2 \rangle_\infty$ for the branching algorithm.}
\end{figure}

Figures~\ref{fig:Ycol_wrt_B_hom_100} and \ref{fig:Ycol_wrt_B_het_60} show the dependence of the spatially-averaged Feynman moments of the collision counts with respect to the bin size $\delta$, respectively in homogeneous and heterogeneous systems. All the Feynman moments of the collision counts scale as a power law of $\delta$, for $\delta \lesssim l_\text{clust}$. Additionally, Fig.~\ref{fig:Ycol_wrt_B_hom_100} saturates around $\sqrt{\langle r^2 \rangle_\infty}$, similarly to the Feynman moment of the fission source. This last remark does not hold for system 4 (see Fig.~\ref{fig:Ycol_wrt_B_het_60}), which is related to the discussion on Fig.~\ref{fig:Yfiss_wrt_B_het_60} (in Section~\ref{subsubsection:fission_feynman}), where we observed the same trend for the dependency on $\delta$ of the Feynman moments for the fission source. In any case, branchless collisions lead to Feynman moments that are considerably smaller than those of branching collisions, and all population control algorithms seem to have little effect on the dependence on coarse-graining of the Feynman moments of the collision counts. The dependence on $N$ and $G$ of the Feynman moment of the collision counts is similar to that of the fission source, and for the sake of conciseness it will not be shown here.

\begin{figure}[t]
    \centering
    \begin{subfigure}{0.45\textwidth}
        \includegraphics[width=\textwidth]{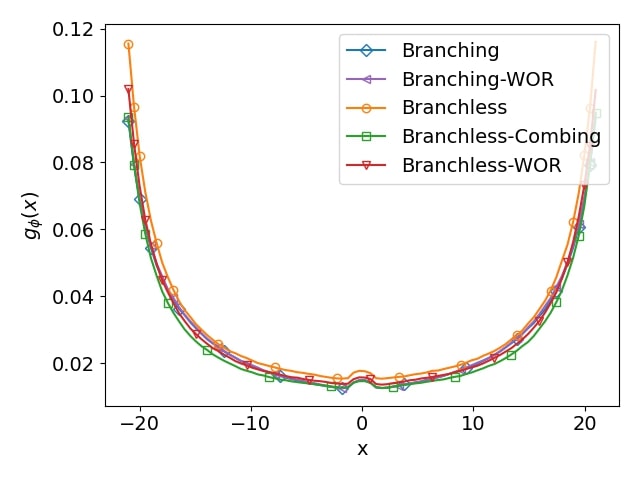}
        \caption{System 3.}
        \label{fig:het_42_gcol}
    \end{subfigure}
    \begin{subfigure}{0.45\textwidth}
        \includegraphics[width=\textwidth]{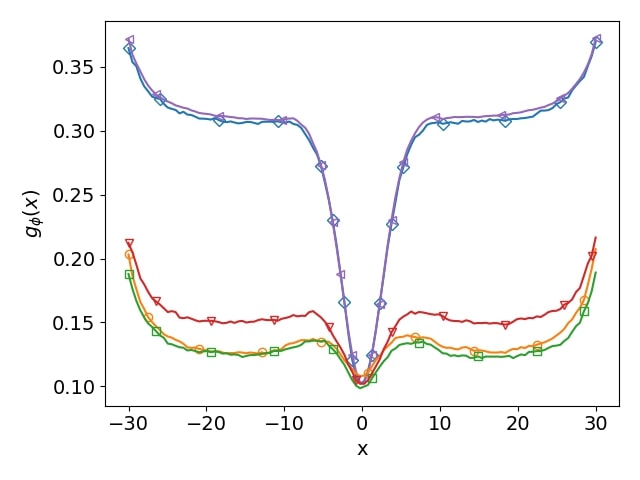}
        \caption{System 4.}
        \label{fig:het_60_gcol}
    \end{subfigure}
    \caption{Equilibrium normalized variance for the collision counts of neutrons, computed for $N=1000$, $\delta=0.5$, and with $10^4$ independent replicas, for the two heterogeneous systems.}
    \label{fig:pi_gcol}
\end{figure}

\subsection{Analysis of the collision counts: normalized variance}

In the homogeneous cases, the normalized variance of the collision counts behaves similarly to that of the fission source (not shown here for conciseness). The two heterogeneous configurations are shown in Fig.~\ref{fig:pi_gcol}. In both cases 3 and 4, it is interesting to note that the peaks observed in the corresponding Feynman moments do not occur in the normalized variance. This illustrates clearly the difference between Feynman moments and normalized variances.

\begin{figure}[t]
    \centering
    \begin{subfigure}{0.45\textwidth}
        \includegraphics[width=\textwidth]{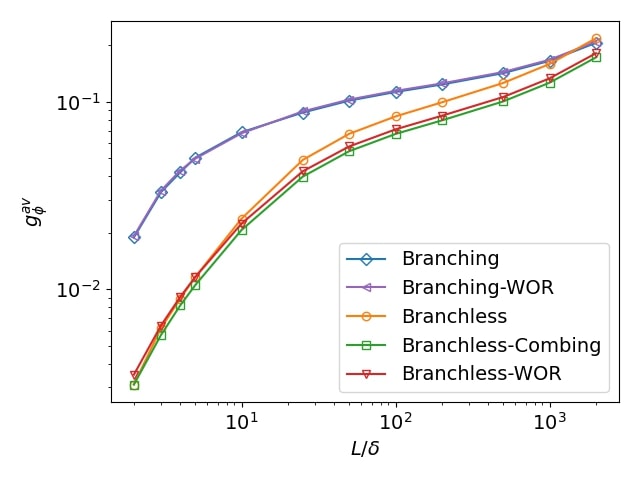}
        \caption{System 1, with $10^5$ independent replicas.}
        \label{fig:hom_50_gcol_B}
    \end{subfigure}
    \begin{subfigure}{0.45\textwidth}
        \includegraphics[width=\textwidth]{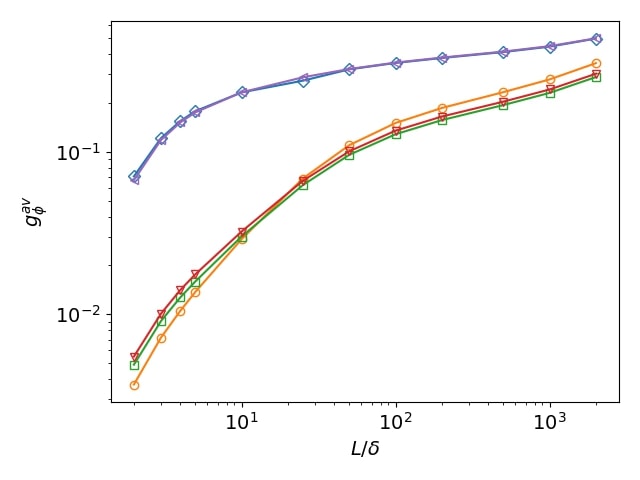}
        \caption{System 2, with $10^5$ independent replicas.}
        \label{fig:hom_100_gcol_B}
    \end{subfigure}
    \begin{subfigure}{0.45\textwidth}
        \includegraphics[width=\textwidth]{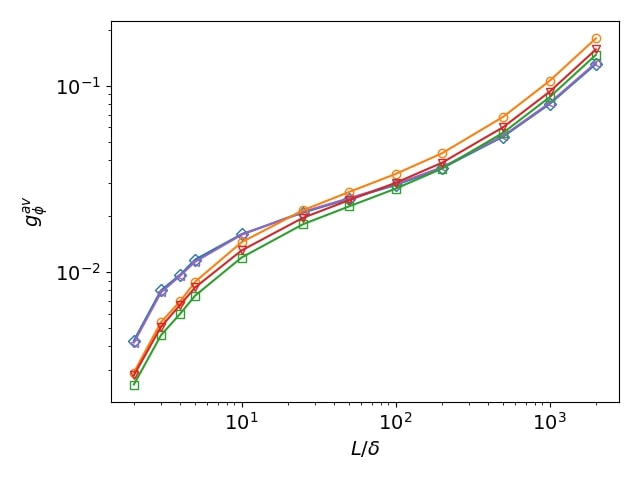}
        \caption{System 3, with $10^5$ independent replicas.}
        \label{fig:het_42_gcol_B}
    \end{subfigure}
    \begin{subfigure}{0.45\textwidth}
        \includegraphics[width=\textwidth]{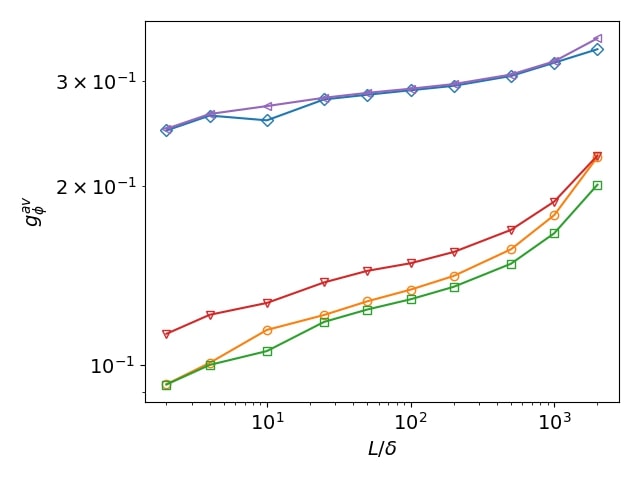}
        \caption{System 4, with $10^4$ independent replicas.}
        \label{fig:het_60_gcol_B}
    \end{subfigure}
    \caption{Space-averaged equilibrium normalized variance for the collision counts of neutrons as a function of bin size, computed for $N=1000$ and $B=100$ for different systems.}
    \label{fig:pi_gcol_B}
\end{figure}

Finally, we consider the dependence of normalized variance with the size of spatial cells, which is expected to scale as $1/\delta$ when $\delta$ is small enough~\cite{zoia_neutron_2017}; indeed, Fig.~\ref{fig:pi_gcol_B} shows that the normalized variance behaves as $1/\delta$. When the dominance ratio is close to unity (see Fig.~\ref{fig:het_60_gcol_B}), the normalized variance for the branchless algorithm is significantly lower than for the branching algorithm. Additionally, although all three branchless algorithms yield similar normalized variances, branchless collisions in combination with combing appears to consistently yield the lowest normalized variance in our systems.

\section{Conclusions}
\label{sec:conclusions}

In this work, we have investigated the impact of sampling strategies, variance reduction and population control methods on the correlations that occur in the Monte Carlo implementation of the power iteration algorithm. For this purpose, we have selected several distinct tallies: the Shannon entropy, the average square pair distance, the average number of surviving families, the Feynman moments and the normalized variance. This analysis has been carried out on a set of simple, yet meaningful one-dimensional slab benchmark configurations, encompassing homogeneous and heterogeneous geometries with several degrees of decoupling. In order to probe the effects of each technique, the power iteration was run with different combinations of branching or branchless collisions, and combing or sampling WOR.

All the proposed tallies help in some way to detect the presence of anomalous fluctuations in the examined systems. However, each comes with specific advantages and drawbacks. Global tallies, such as the entropy and the average square pair distance, while being easy to use, have been shown to be inadequate for the investigation of heterogeneous systems, due to lack of information on the spatial details. Conversely, local (i.e.\ space dependent) tallies such as the Feynman moments and the normalized variance are inherently useful in extracting information on the spatial behavior of the correlations, at the expense of an increased complexity.

The main finding of our work is that, in all tested configurations, the use of branchless collisions (as opposed to regular branching collisions) is very effective at quenching correlations. In particular, the use of branchless collisions makes the apparent (ergodic) variance much closer to the real (ensemble) variance. In addition, while population control is generally advantageous in terms of weakening the effects of correlations, the impact of these methods is much milder than that of using branchless collisions. Sampling without replacement was showed to lose its effectiveness in a multi-group framework with non-analog collisions, as opposed to the single-speed, analog configurations where it was originally tested.

Future work will focus on extending the findings discussed in this paper in two directions. On the one hand, we will address a broader ensemble of benchmark configurations, and assess in particular whether our claims hold true in more realistic three-dimensional reactor models, especially using the continuous-energy representation of nuclear data. On the other hand, we will examine whether the benefits of branchless collision sampling carry over to more complex simulations where power iteration is coupled to thermal-hydraulics or depletion solvers.

\section*{Acknowledgments}
The authors wish to thank Dr.~F.~Malvagi (IRSN) for sharing unpublished simulation results.

\pagebreak
\bibliographystyle{style/ans_js} 
\bibliography{main}

\end{document}